\documentclass[twocolumn]{aastex631}  

\usepackage{graphicx}
\usepackage{txfonts}
\usepackage{hyperref}
\hypersetup{
	colorlinks,
	citecolor=blue,
	filecolor=black,
	linkcolor=black,
	urlcolor=black
}
\usepackage{natbib}
\begin{document}

\title{Precise Transit Photometry Using TESS II: Revisiting 28 Additional Transiting Systems With Updated Physical Properties}

\author[0000-0001-8018-0264]{Suman Saha}
\affiliation{Instituto de Estudios Astrofísicos, Facultad de Ingeniería Ciencias, Universidad Diego Portales, Av. Ejército Libertador 441, Santiago, Chile}
	
\correspondingauthor{Suman Saha}
\email{suman.saha@mail.udp.cl}

\accepted{for publication in The Astrophysical Journal Supplement Series}

\begin{abstract}
Precise physical properties of the known transiting exoplanets are essential for their precise atmospheric characterization using modern and upcoming instruments. Leveraging the large volume of high SNR photometric follow-up data from TESS, highly precise physical properties can be estimated for these systems, especially for those discovered using ground-based instruments prior to the TESS mission. In this work, I have used the publicly available TESS follow-up data for 28 transiting systems with 10 $<$ V$_{mag}$ $<$ 10.5, with an aim to update their known physical properties. The observed lightcurves have been analysed by implementing a state-of-the-art critical noise treatment algorithm to effectively reduce both time-correlated and un-correlated noise components, using sophisticated techniques like wavelet denoising and Gaussian-process regression. Compared with the previous studies, the estimated transit parameters are found to be more precise for most of the targets, including a few cases where a larger space-based instrument like Spitzer, Kepler or CHEOPS has been used in the previous study. The large volume of transit observations used for each target has also resulted in a more accurate estimation of the physical properties, as this overcomes any error in parameter estimations from bias present in a smaller volume of data. Thus, comparing with the literature values, statistically significant improvements in the known physical properties of several targeted systems have been reported from this work. The large volume of transit timing information from the analyses was also used to search for Transit Timing Variation trends in these targets, which has resulted in no significant detection.
\end{abstract}


%

\section{Introduction}\label{sec:sec1}

The Transiting Exoplanet Survey Satellite \citep[TESS,][]{2015JATIS...1a4003R} is the largest ongoing survey mission for the detection of transiting exoplanets around bright sources in the Solar neighbourhood. Having a very large field of view (24×96 degrees, also known as sectors), TESS will survey over almost the entirety of the sky in several sectors. This allows TESS to survey a large number of stars over a long time-scale, which is potentially going to result in the detection of thousands of transiting exoplanets around bright stars. This also allows TESS to automatically obtain a large volume of follow-up photometric data for a large number of known systems.

\begin{table*}[]
    \centering
    \caption{Targets and observational details}
    \label{tab:tab1}
    $\begin{array}{lccc}
        \hline
        \hline
         \text{Target Name} & \text{Host Star $V_{mag}$} & \text{Sector} & \text{No. of full Transits} \\
         \hline
         \text{WASP-79 b} & 10.044 & 4,5,31,32 & 24 \\
         \text{WASP-94A b} & 10.051 & 1, 27, 68 & 18 \\
         \text{WASP-131 b} & 10.068 & 11, 64 & 9 \\
         \text{WASP-82 b} & 10.073 & 5 & 10 \\
         \text{HAT-P-67 b} & 10.106 & 24, 26, 51-53 & 22 \\
         \text{WASP-117 b} & 10.139 & 2, 3, 29, 30, 69 & 11\\
         \text{WASP-127 b} & 10.148 & 9, 35, 46, 62, 72 & 23\\
         \text{KELT-18 b} & 10.16 & 15, 16, 22, 23, 48, 50 & 18\\
         \text{HAT-P-49 b} & 10.205 & 14, 41, 55 & 27\\
         \text{WASP-62 b} & 10.213 & 1-4, 6-13, 27-34, 36-39, 61, 62, 64, 65, 68, 69 & 153\\
         \text{XO-6 b} & 10.247 & 19, 20, 26, 40, 53, 59, 60 & 40\\
         \text{WASP-34 b} & 10.285 & 9, 36, 63 & 14 \\
         \text{WASP-77A b} & 10.294 & 4, 31 & 31\\
         \text{WASP-187 b} & 10.295 & 17, 57 & 8\\
         \text{KELT-23A b} & 10.308 & 14-17, 21, 23, 41, 47-50, 54, 57 & 119\\
         \text{WASP-101 b} & 10.336 & 6, 33 & 12\\
         \text{HAT-P-30 b} & 10.352 & 7, 34, 61 & 24\\
         \text{HAT-P-8 b} & 10.358 & 56 & 8\\
         \text{KELT-6 b} & 10.371 & 23, 49 & 4\\
         \text{HAT-P-17 b} & 10.375 & 15, 55, 56 & 6\\
         \text{HAT-P-34 b} & 10.403 & 14, 41, 54, 55 & 13\\
         \text{WASP-54 b} & 10.413 & 23, 46, 50 & 13\\
         \text{HAT-P-13 b} & 10.421 & 47 & 8\\
         \text{WASP-13 b} & 10.424 & 21 & 7\\
         \text{WASP-73 b} & 10.468 & 1, 27, 67 & 16\\
         \text{HAT-P-6 b} & 10.469 & 16, 17, 57 & 16\\
         \text{HAT-P-7 b} & 10.481 & 14, 15, 40, 41, 54, 55 & 66\\
         \text{KELT-21 b} & 10.484 & 14, 15, 41, 55 & 12\\
         \hline
    \end{array}$
\end{table*}

Being a space-based instrument, the observations from TESS are not affected by the variability and perturbations of the Earth's atmosphere, which severely affects the ground-based observations. The ground-based observations are also affected by day-night cycles and climatic conditions. Most of the ground-based instruments which are dedicated for the detection of transiting exoplanets are comparatively small, and combined with the perturbations due to Earth's atmosphere, the sensitivity of these instruments is pretty low. On the other hand, the larger ground-based instruments are limited in number and share their time among various astronomical studies, and thus can not be used for large-scale follow-up observations of the known exoplanets. All these factors combine to make TESS a very interesting and unique facility presently available for such large-scale follow-up studies.

As a highly sophisticated space-bound telescope, the photometric observations from TESS have very high SNR, comparable to much larger ground-based telescopes for bright sources. Besides, due to the long uninterrupted time-series observations obtained from TESS over each sector and there being a large probability that each exoplanet hosting system would be observable over multiple TESS sectors, large volumes of photometric data covering several transit events could be obtained for each such systems. Such large volumes of transit photometric data are useful for high-precision photometric studies, making it possible to estimate the physical properties of the transiting systems with a very high degree of accuracy and precision. Such precise estimation of the planetary parameters is extremely useful for high-precision spectroscopic studies using the most modern and sophisticated instruments like HST, JWST, VLT etc., as well as the upcoming instruments like GMT and ELT \cite[e.g. for WASP-79 b:][]{2017AJ....153..136S, 2020AJ....160..109S, 2022MNRAS.514.5192L, 2022AJ....163....7F}. Since the planetary systems around comparatively bright sources are deemed to be the prime targets for such precise atmospheric characterization, it would be prudent to use the TESS follow-up observations for such targets to precisely estimate their physical properties. The large volume of data from TESS would also be instrumental in removing bias in the estimated properties from a smaller volume of data, which is true for several of the exoplanets detected around the bright sources using ground-based observations previously. This could result in statistically significant improvements in the known properties of these planets, which is significant for adopting these parameters for future spectroscopic follow-up studies. A homogeneously derived highly precise and accurate set of properties for a large number of known exoplanets, especially those around the brighter sources, would also be useful for studies involving statistical analyses of various planetary populations \cite[e.g.][]{2016ApJ...831...64T, 2019A&A...630A.135U, 2020ApJ...891...12N}. The highly accurate ephemerides from the large number of transits detected from TESS can be combined with those from the previous observations in order to obtain highly precise ephemeris estimations for several known exoplanets, which will be useful for future follow-up studies over a longer time period \cite[e.g.][]{2022ApJS..259...62I, 2022ApJS..258...40K}. Finally, the large volume of transit timing information for these exoplanets from the modelling of these transit follow-up data would also be useful to search for any Transit Timing Variation (TTV) trends in these systems to study the extreme dynamics like orbital decay/enhancements \citep[e.g.][]{2020AJ....159..150P, 2022AJ....163..175W, 2024ApJS..270...14W}, precession \citep[e.g.][]{2008ApJ...685..543J, 2010ApJ...716..850C}, as well as the search for undetected exoplanets \citep[e.g.][]{2011ApJS..197....7C, 2021AJ....161..161D} and exomoons \citep[e.g.][]{2009MNRAS.392..181K, 2021MNRAS.501.2378F, 2022ApJ...936....2S, 2024BSRSL..93..123S}.

In my previous work \citep[i.e.][]{2023ApJS..268....2S}, the TESS follow-up observations for 28 exoplanets around very bright sources with V$_{mag}$ $<$ 10 have been analysed, which has resulted in a set of more accurate and precise estimations of the physical properties for those systems. The analysis involved a critical noise treatment algorithm leveraging sophisticated techniques like wavelet denoising and Gaussian process regression \citep[e.g.][]{2021AJ....162...18S, 2021AJ....162..221S, saha2022precise, 2023ApJS..268....2S}, such that noise components both uncorrelated and correlated in time could be reduced efficiently from the lightcurves.Both wavelet denoising and Gaussian process regression techniques have been extensively tested in the previous studies to determine their relative impact in reducing noise levels in the lightcurves and lowering the uncertainties in the estimated parameters \citep{2019AJ....158...39C, 2021AJ....162..221S}. In this work, I have analysed the archival TESS follow-up data for 28 additional exoplanets with 10 $<$ V$_{mag}$ $<$ 10.5, with the aim of improving the accuracy and precision of their known physical properties compared to the best-known literature values. A critical noise treatment algorithm similar to \cite{2023ApJS..268....2S} has been used to treat various noise components, such as stellar variabilities and instrumental systematics, and a comparative analysis of the estimated parameters with the literature values has been presented.

Section \ref{sec:sec2} details the target selection and observational data used in this work. In section \ref{sec:sec3}, the data analysis and modelling techniques used for analyses have been explained. And finally, in section \ref{sec:sec4}, the results and outcomes from the study have been discussed.

\begin{figure*}
	\centering
	\includegraphics[width=0.94\linewidth]{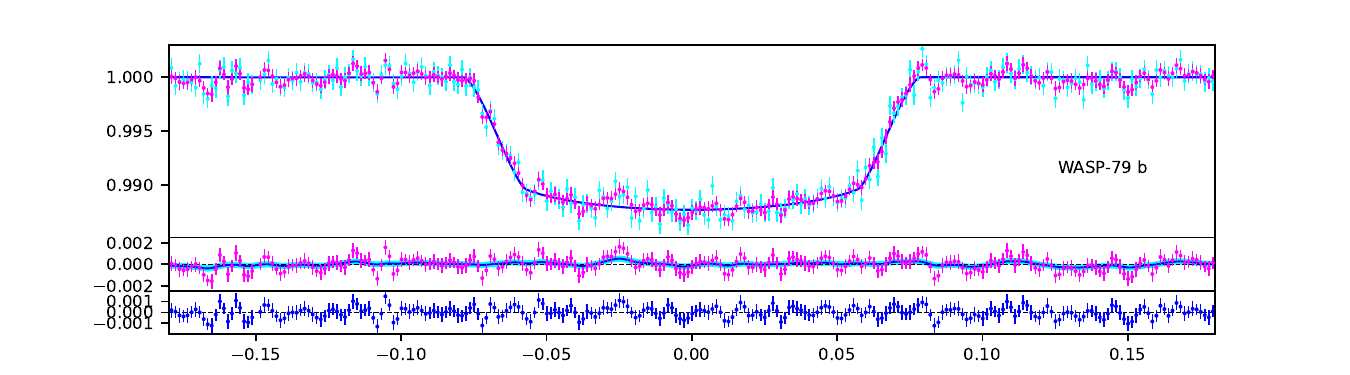}
	\includegraphics[width=0.94\linewidth]{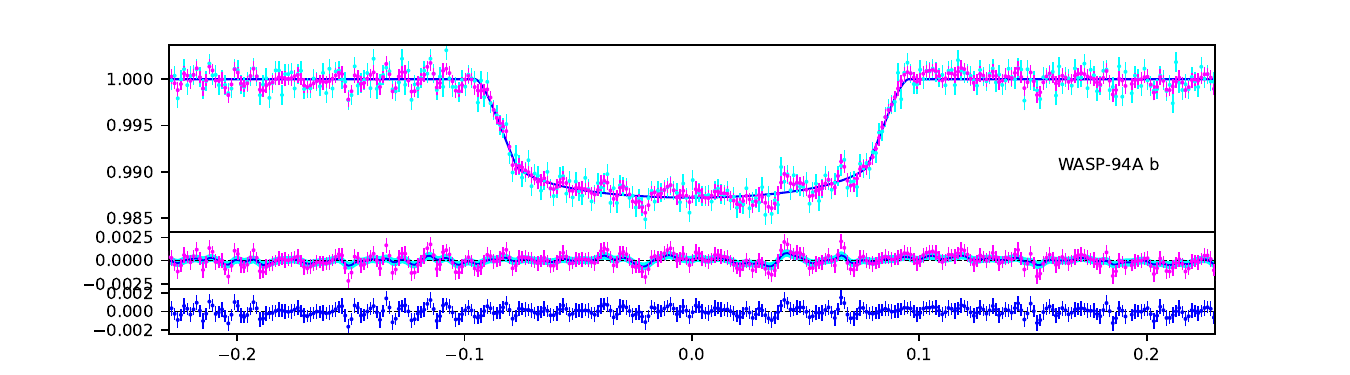}
	\includegraphics[width=0.94\linewidth]{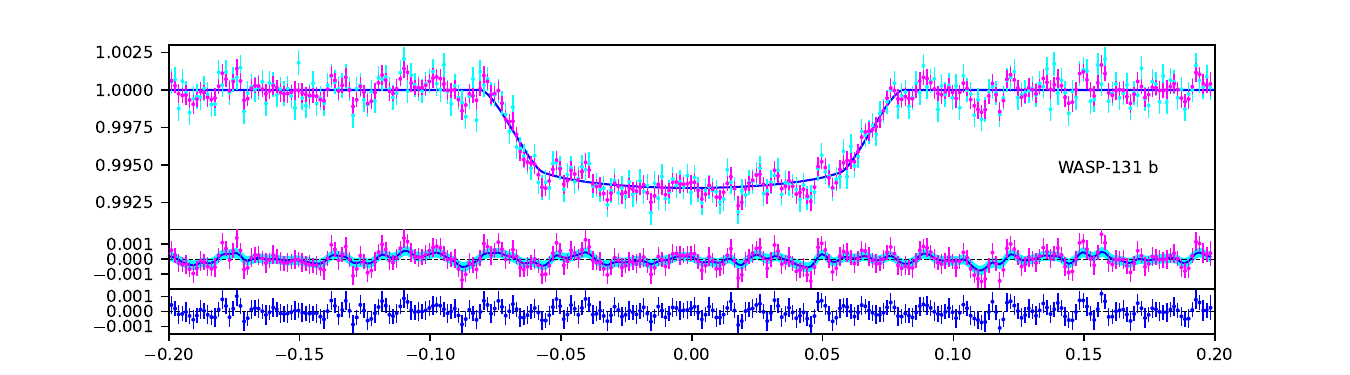}
	\includegraphics[width=0.94\linewidth]{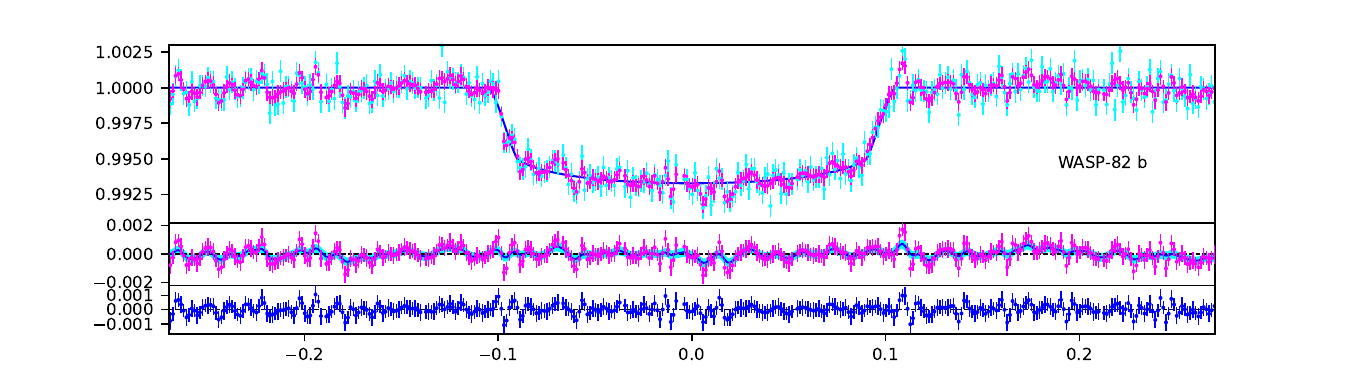}
	\includegraphics[width=0.94\linewidth]{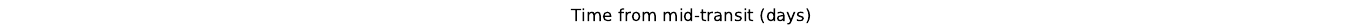}
	\caption{Observed and best-fit model light-curves (one transit event) for WASP-79 b, WASP-94A b, WASP-131 b and WASP-82 b. For each observed transit, Top: the unprocessed light-curve (cyan), light-curve after wavelet denoising (magenta), the best-fit transit model (blue). Middle: the residual after modelling without GP regression (magenta), the mean (blue) and 1-$\sigma$ interval (cyan) of the best-fit GP regression model. Bottom: mean residual flux (blue). The mean residual flux corresponds to the residual flux considering the mean of the best fit GP regression model.}
	\label{fig:fig1}
\end{figure*}

\begin{figure*}
	\centering
	\includegraphics[width=0.94\linewidth]{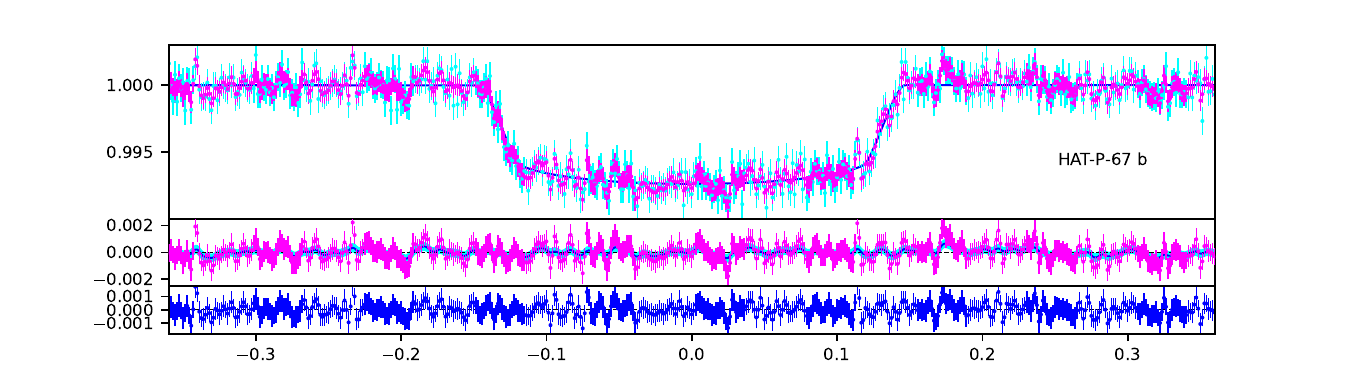}
	\includegraphics[width=0.94\linewidth]{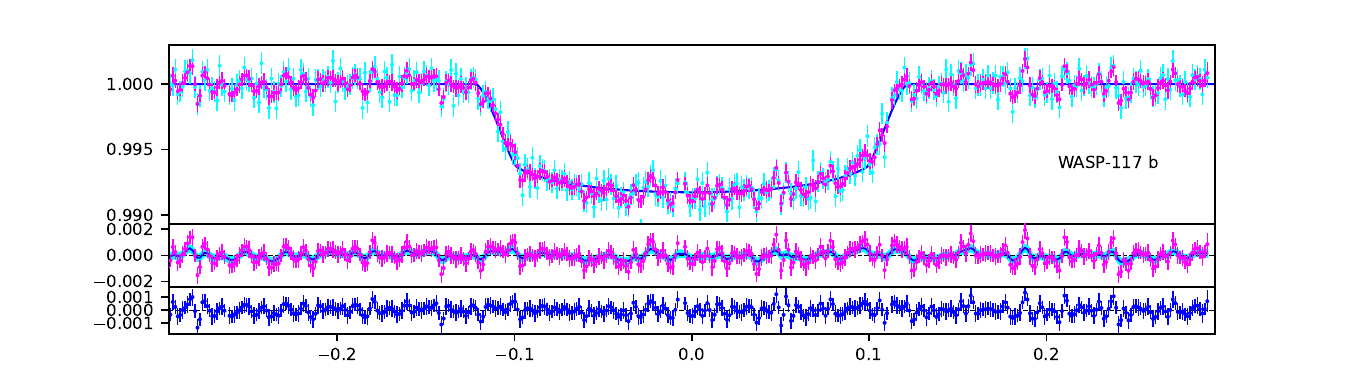}
	\includegraphics[width=0.94\linewidth]{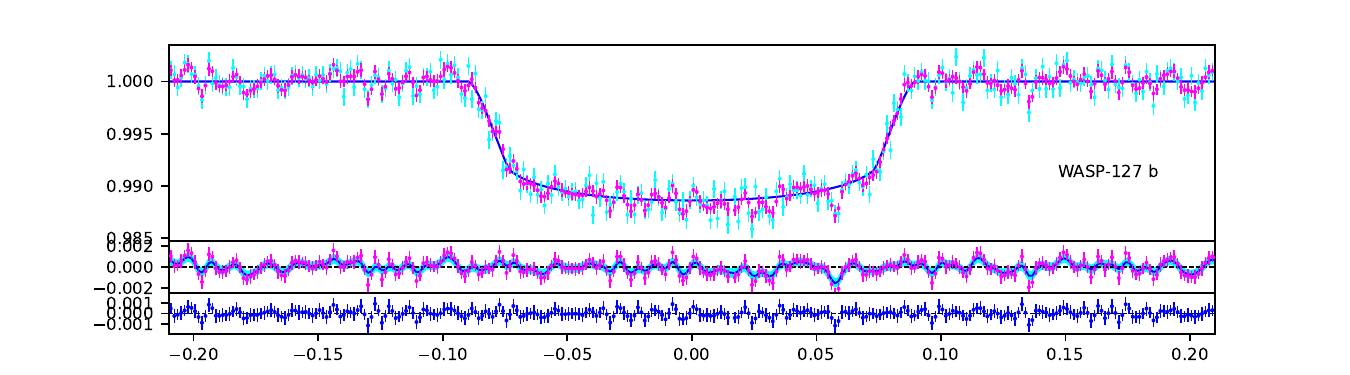}
	\includegraphics[width=0.94\linewidth]{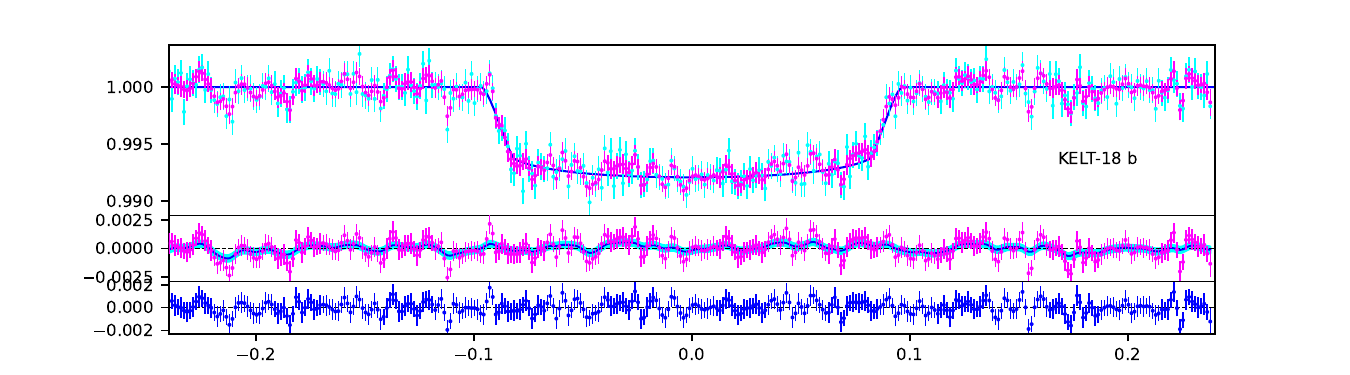}
    \includegraphics[width=0.94\linewidth]{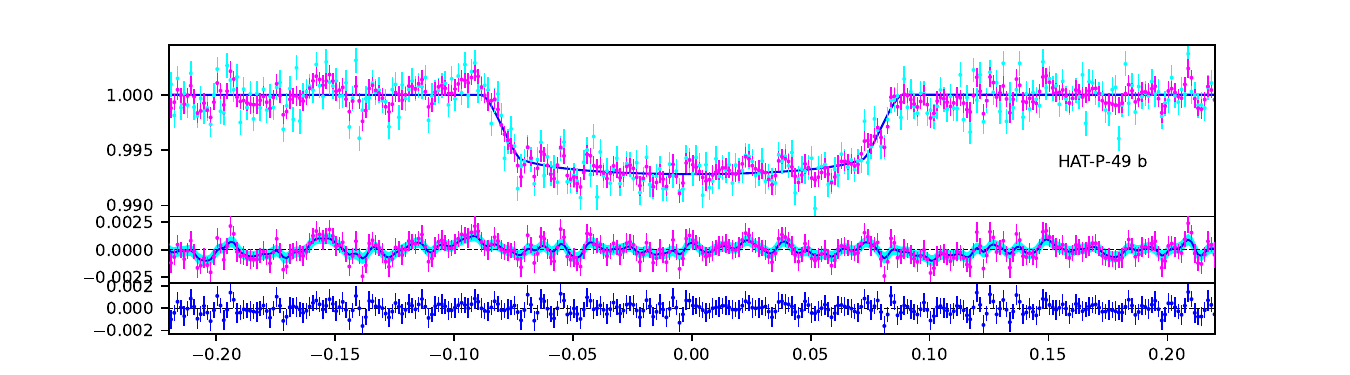}
	\includegraphics[width=0.94\linewidth]{f0.pdf}
	\caption{Same as Figure \ref{fig:fig1}, but for HAT-p-67 b, WASP-117 b, WASP-127 b, KELT-18 b and HAT-P-49 b.}
	\label{fig:fig2}
\end{figure*}

\begin{figure*}
	\centering
	\includegraphics[width=0.94\linewidth]{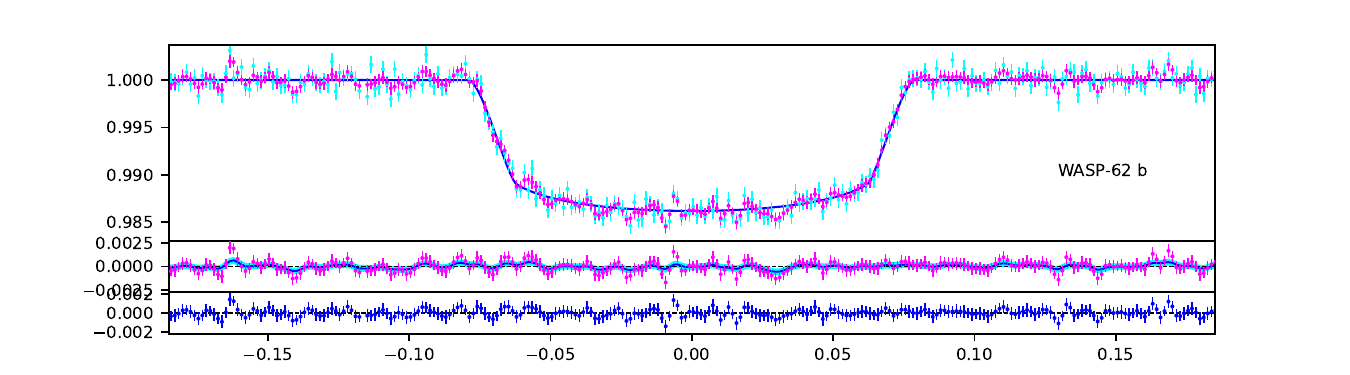}
	\includegraphics[width=0.94\linewidth]{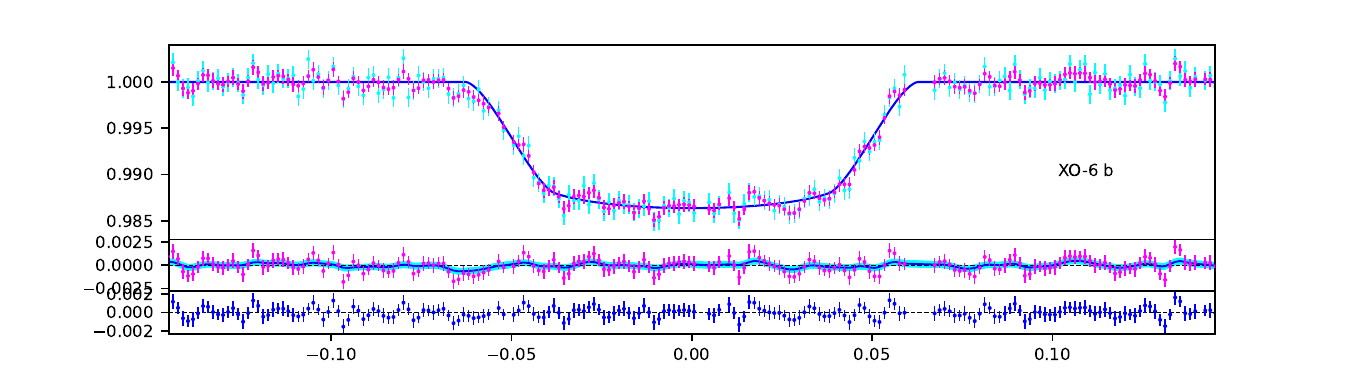}
	\includegraphics[width=0.94\linewidth]{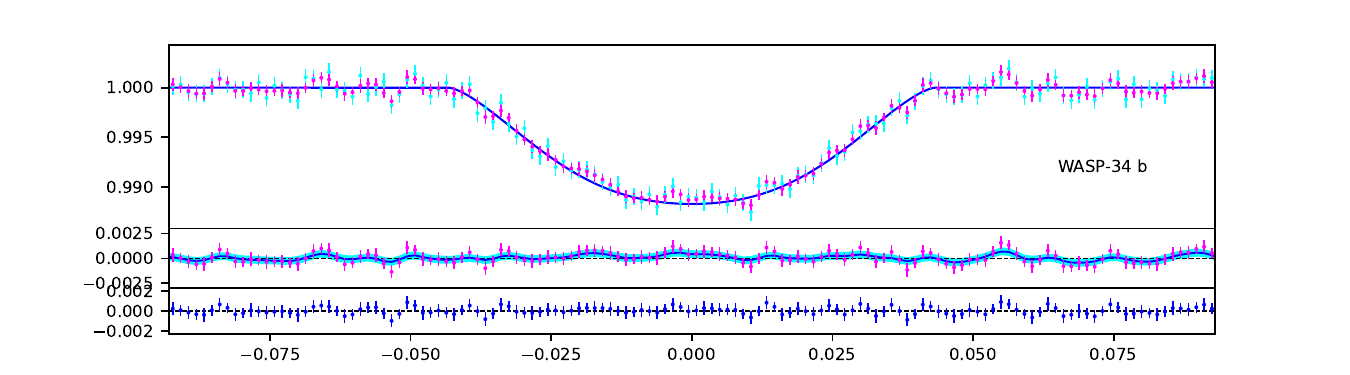}
	\includegraphics[width=0.94\linewidth]{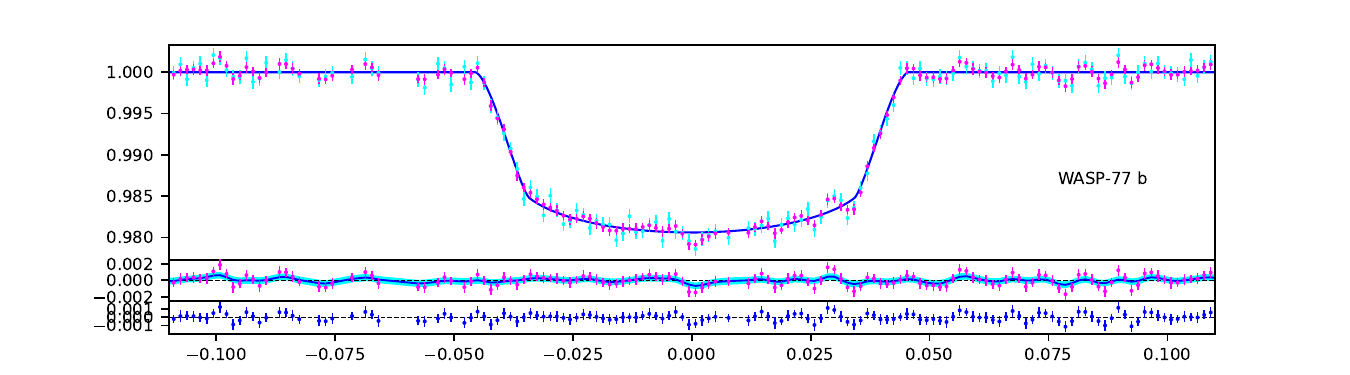}
    \includegraphics[width=0.94\linewidth]{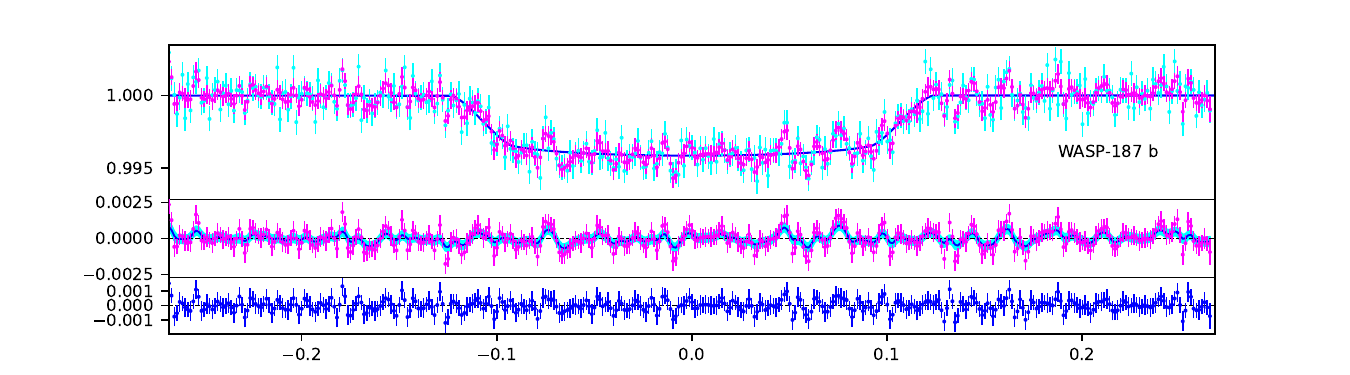}
	\includegraphics[width=0.94\linewidth]{f0.pdf}
	\caption{Same as Figure \ref{fig:fig1}, but for WASP-62 b, XO-6 b, WASP-34 b, WASP-77 b and WASP-187 b.}
	\label{fig:fig3}
\end{figure*}

\begin{figure*}
	\centering
	\includegraphics[width=0.94\linewidth]{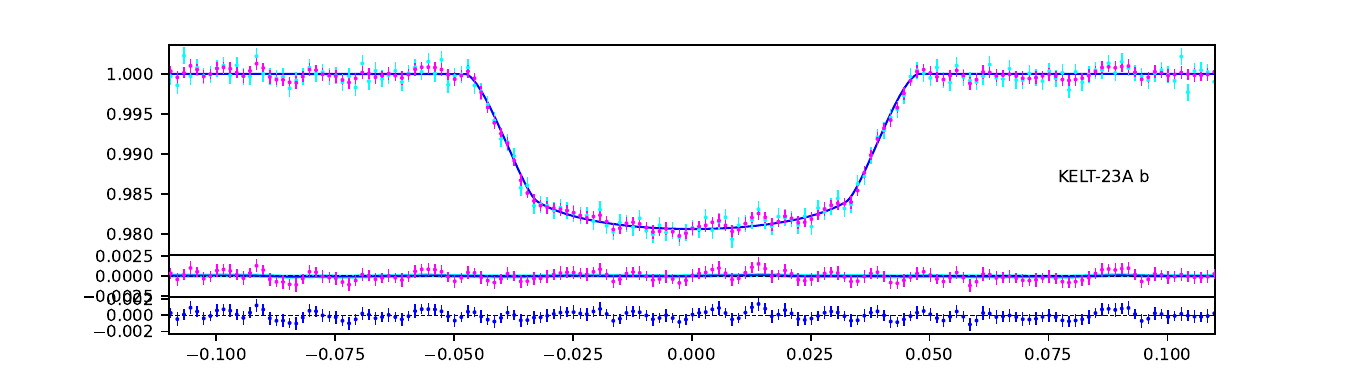}
	\includegraphics[width=0.94\linewidth]{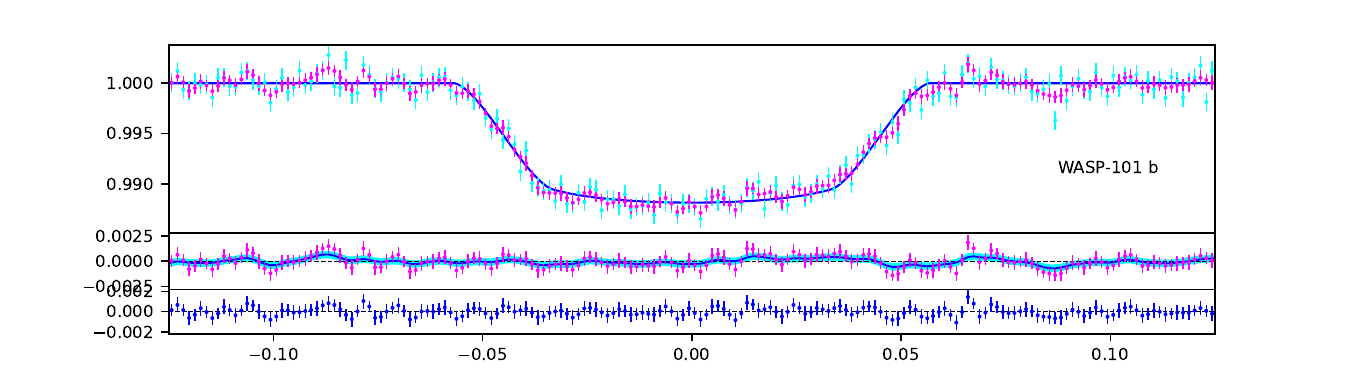}
	\includegraphics[width=0.94\linewidth]{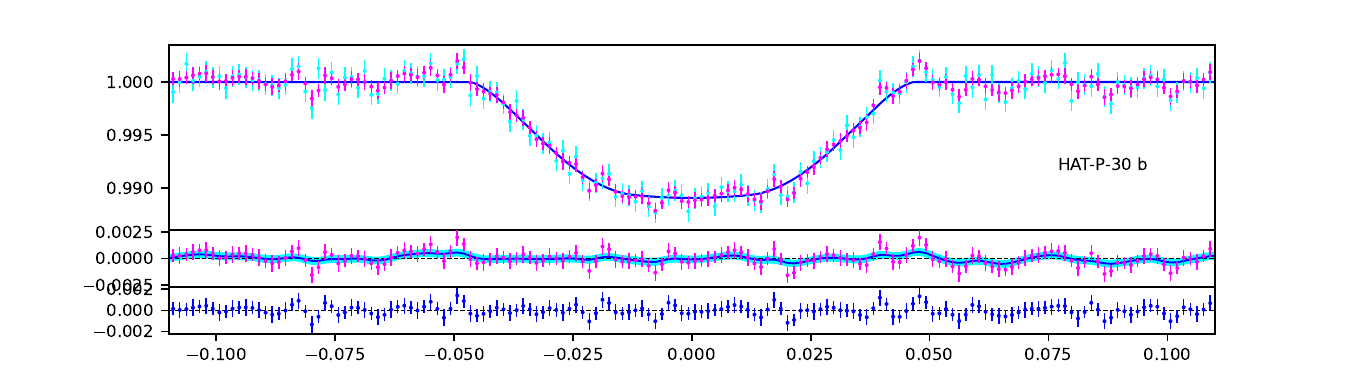}
	\includegraphics[width=0.94\linewidth]{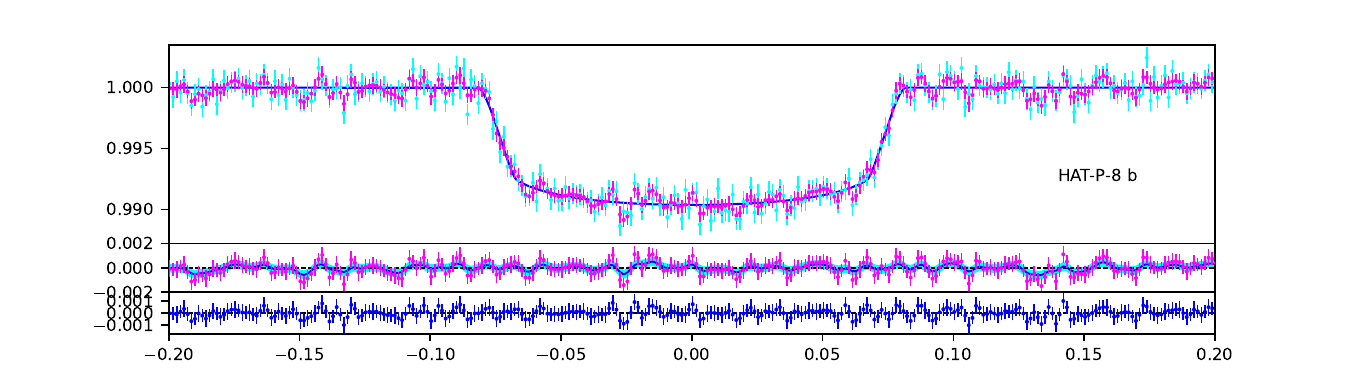}
    \includegraphics[width=0.94\linewidth]{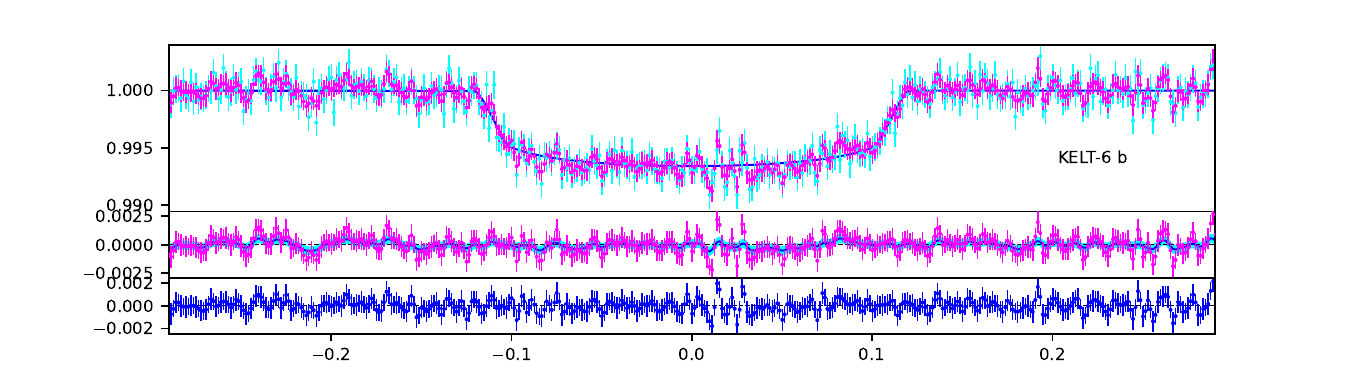}
	\includegraphics[width=0.94\linewidth]{f0.pdf}
	\caption{Same as Figure \ref{fig:fig1}, but for KELT-23A b, WASP-101 b, HAT-P-30 b, HAT-P-8 b and KELT-6 b.}
	\label{fig:fig4}
\end{figure*}

\begin{figure*}
	\centering
	\includegraphics[width=0.94\linewidth]{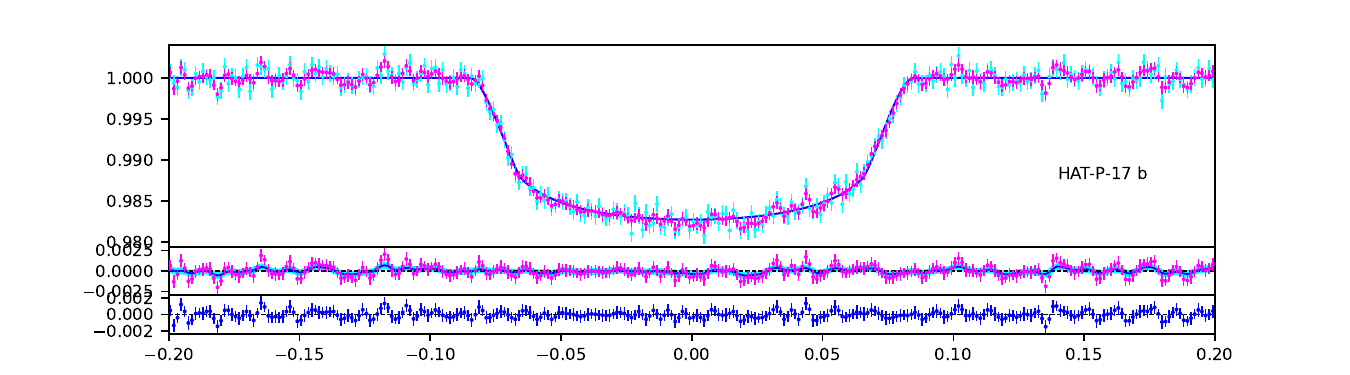}
	\includegraphics[width=0.94\linewidth]{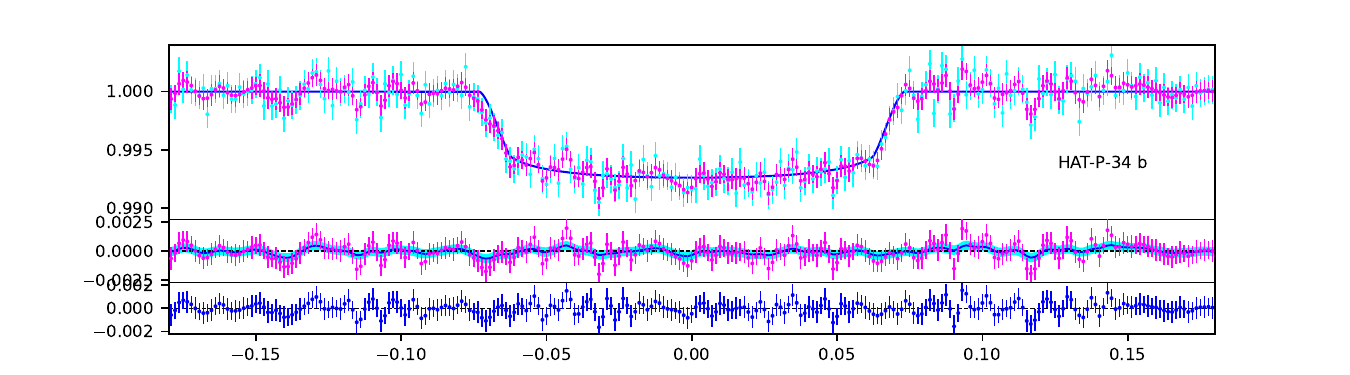}
	\includegraphics[width=0.94\linewidth]{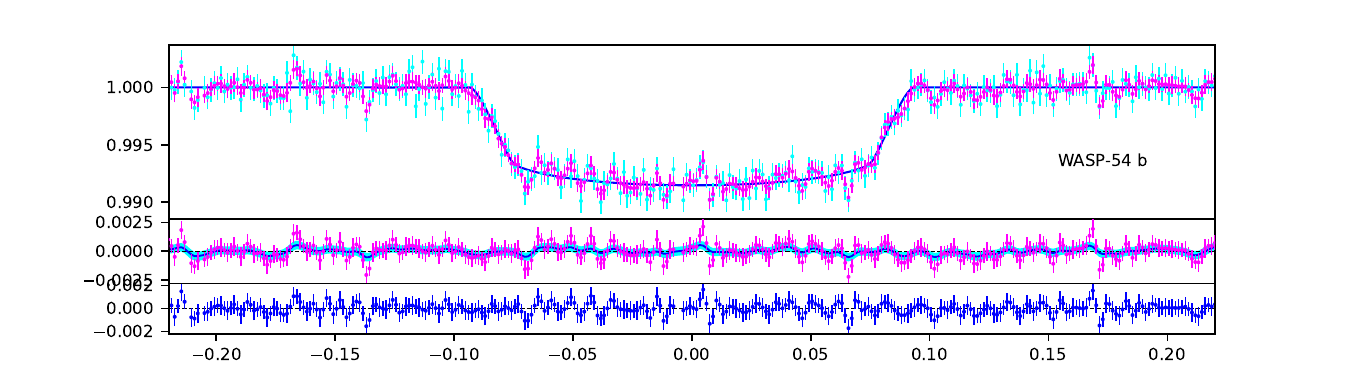}
	\includegraphics[width=0.94\linewidth]{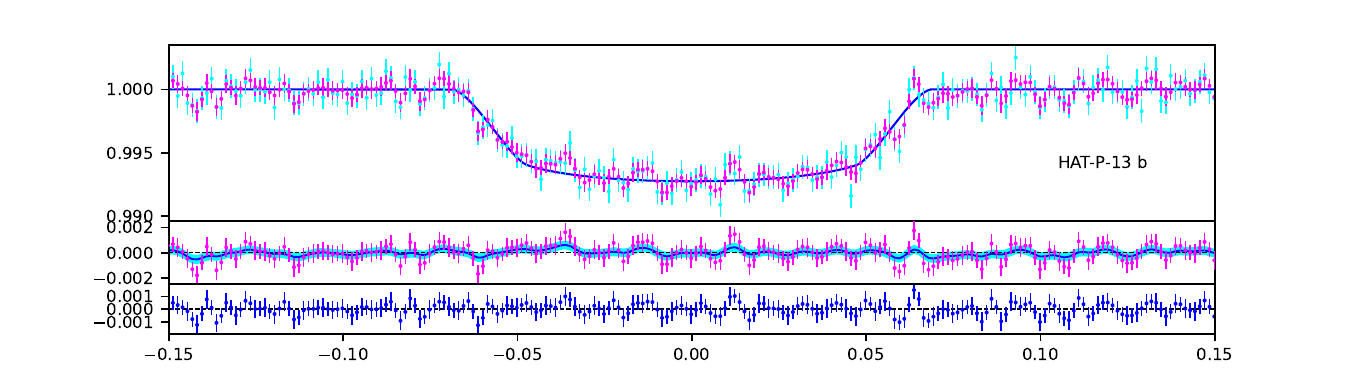}
    \includegraphics[width=0.94\linewidth]{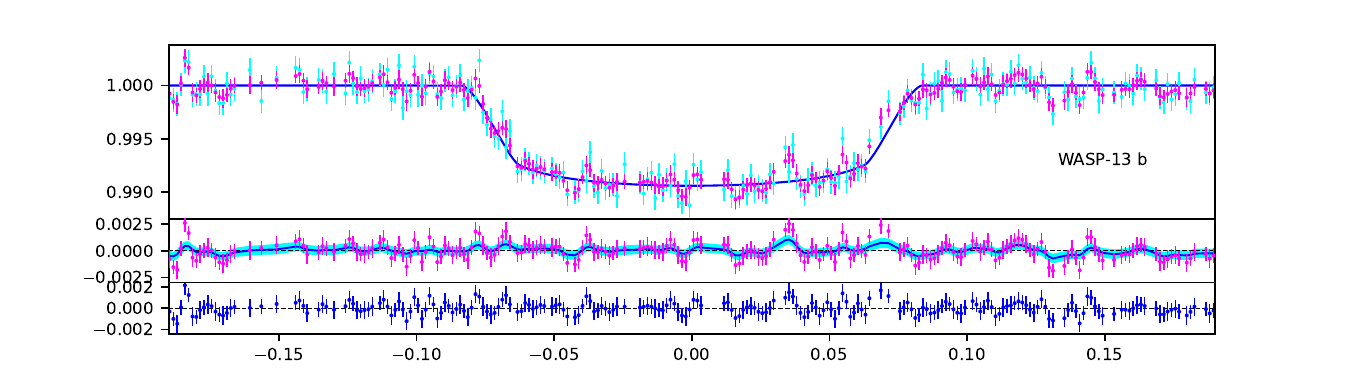}
	\includegraphics[width=0.94\linewidth]{f0.pdf}
	\caption{Same as Figure \ref{fig:fig1}, but for HAT-P-17 b, HAT-P-34 b, WASP-54 b, HAT-P-13 b and WASP-13b.}
	\label{fig:fig5}
\end{figure*}

\begin{figure*}
	\centering
	\includegraphics[width=0.94\linewidth]{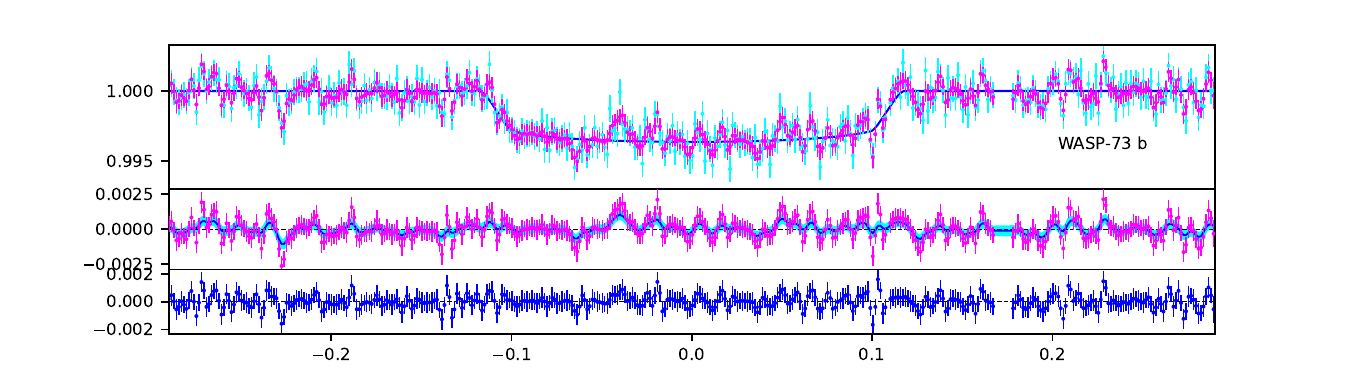}
	\includegraphics[width=0.94\linewidth]{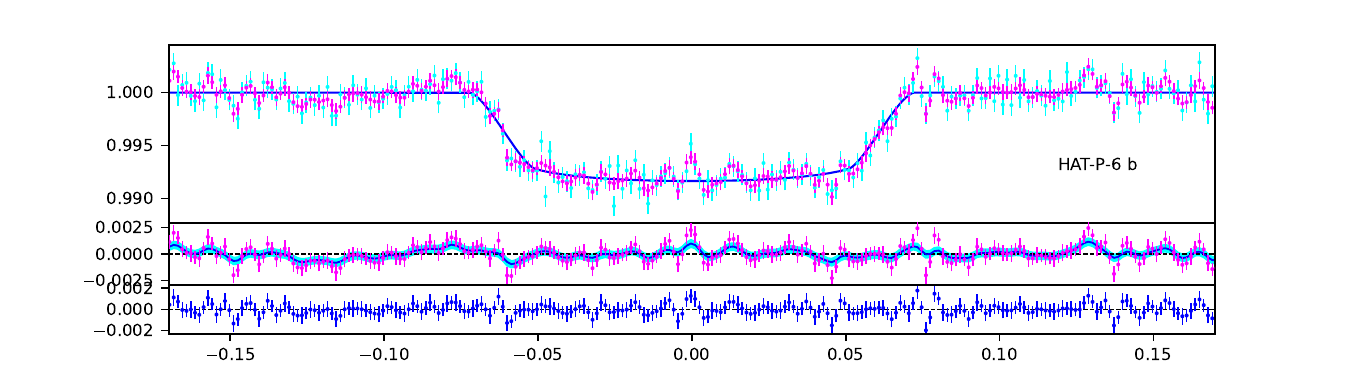}
	\includegraphics[width=0.94\linewidth]{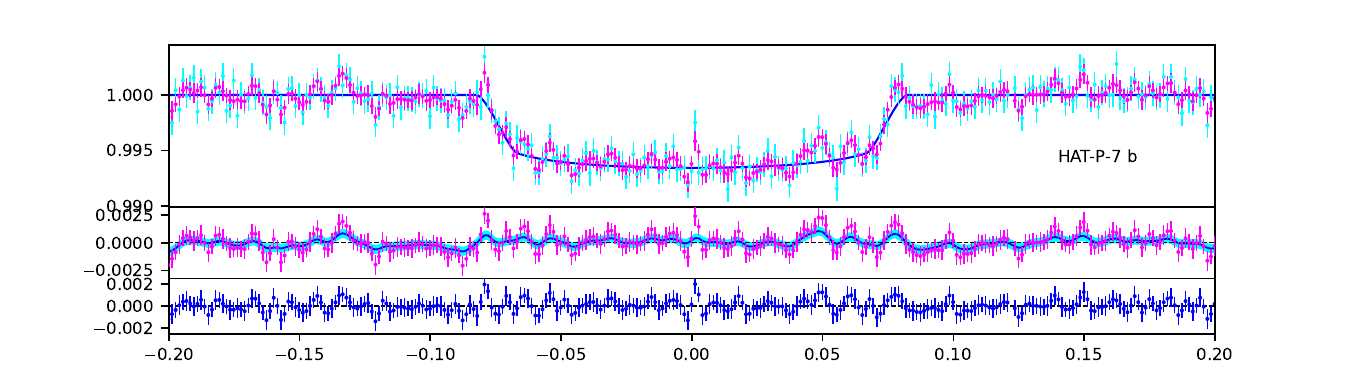}
	\includegraphics[width=0.94\linewidth]{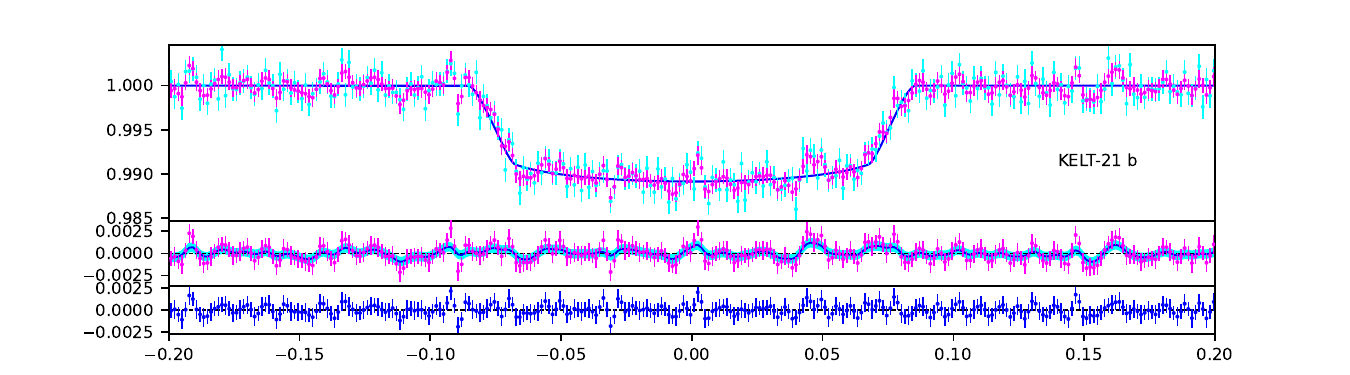}
	\includegraphics[width=0.94\linewidth]{f0.pdf}
	\caption{Same as Figure \ref{fig:fig1}, but for WASP-73 b, HAT-P-6 b, HAT-P-7 b and KELT-21 b.}
	\label{fig:fig6}
\end{figure*}

\section{Target selection and observational data}\label{sec:sec2}

This study aims at extending the previous work of updating the physical properties of known exoplanets around bright sources using TESS follow-up observations \citep[i.e.][]{2023ApJS..268....2S}. As the previous study had involved targets around sources with V$_{mag}$ $<$ 10, targets beyond that limit were surveyed for this study, with the criteria being ground-based discovery prior to the TESS mission and the availability of TESS follow-up observations. The NASA Exoplanet Archive\footnote{https://exoplanetarchive.ipac.caltech.edu/} was used to access the list of all known exoplanets, and the Barbara A. Mikulski Archive for Space Telescopes (MAST)\footnote{https://mast.stsci.edu} was used to check for publicly available TESS data for these systems, which have resulted in the selection of 28 targets with the brightest sources (10 $<$ V$_{mag}$ $<$ 10.5), i.e. WASP-79 b, WASP-94A b, WASP-131 b, WASP-82 b, HAT-P-67 b, WASP-117 b, WASP-127 b, KELT-18 b, HAT-P-49 b, WASP-62 b, XO-6 b, WASP-34 b, WASP-77A b, WASP-187 b, KELT-23A b, WASP-101 b, HAT-P-30 b, HAT-P-8 b, KELT-6 b, HAT-P-17 b, HAT-P-34 b, WASP-54 b, HAT-P-13 b, WASP-13 b, WASP73 b, HAT-P-6 b, HAT-P-7 b and KELT-21 b. The corresponding TESS sectors that have observed each of these targets have been listed in Table \ref{tab:tab1}.

The MAST database was used to access the publicly available data for the targets from TESS. Since all of the targets are previously known exoplanets around bright stars, the high cadence (120s) TESS time-series data processed using the Science Processing Operations Center \cite[SPOC,][]{2016SPIE.9913E..3EJ, 2020RNAAS...4..201C} pipeline was available for most of the observed sectors for these targets, which were used in the analyses. Only for a few cases where the data processed by the SPOC pipeline were not available, the data processed by the Quick-Look Pipeline \cite[QLP,][]{2020RNAAS...4..204H} were used instead. A total of 732 full transit events were identified in the lightcurves corresponding to these targets, which have been listed in Table \ref{tab:tab1}. Only the full transit observations have been used in this study to avoid the risk of inaccurate baseline estimations for the partial transit observations, which could affect the accuracy of the estimated parameters.

\begin{table*}[t!]
    \centering
    \caption{Estimated physical parameters for WASP-79 b, WASP-94A b, WASP-131 b and WASP-82 b}
    \label{tab:tab2}
    $\begin{array}{lcccc}
        \hline
        \hline
         \text{Parameter} & \text{WASP-79 b} & \text{WASP-94A b} & \text{WASP-131 b} & \text{WASP-82 b} \\
         \hline
         \text{Transit Parameters} \\
         T_0\; [BJD_{TDB}] & 2458412.891995_{-0.00009}^{+0.000089} & 2458328.29891\pm0.00014 & 2458601.58015\pm0.00031 & 2458438.96647_{-0.00036}^{+0.00035}\\
         P\; [days] & 3.66239094\pm0.0000006 & 3.95020219\pm0.00000049 & 5.3220096\pm0.0000015 & 2.705789_{-0.000066}^{+0.000067}\\
         b & 0.5573_{-0.0095}^{+0.0091} & 0.233_{-0.054}^{+0.051} & 0.724_{-0.018}^{+0.015} & 0.112_{-0.075}^{+0.088}\\
         R_\star/a & 0.14006_{-0.00091}^{+0.00092} & 0.1394_{-0.0014}^{+0.0018} & 0.1184_{-0.0027}^{+0.0026} & 0.2258_{-0.0015}^{+0.0028}\\
         R_p/R_\star & 0.10749\pm0.00025 & 0.10561_{-0.00045}^{+0.00048} & 0.07998_{-0.00061}^{+0.00048} & 0.07683\pm0.00034\\
         \text{Limb-darkening coefficients}\\
         C_1 & 0.321_{-0.041}^{+0.026} & 0.321_{-0.042}^{+0.029} & 0.26_{-0.13}^{+0.11} & 0.306_{-0.065}^{+0.05}\\
         C_2 & 0.06_{-0.037}^{+0.064} & 0.178_{-0.077}^{+0.088} & 0.21_{-0.13}^{+0.17} & 0.134_{-0.079}^{+0.122}\\
         \text{Derived parameters}\\
         T_{14}\; [hr] & 3.77343_{-0.0071}^{+0.007} & 4.566_{-0.011}^{+0.012} & 3.875\pm0.026 & 5.046_{-0.015}^{+0.017}\\
         a/R_\star & 7.139\pm0.047 & 7.173_{-0.089}^{+0.075} & 8.45_{-0.18}^{+0.2} & 4.429_{-0.054}^{+0.029}\\
         i\; [deg] & 85.52\pm0.1 & 88.14_{-0.044}^{+0.045} & 85.08_{-0.21}^{+0.23} & 88.54_{-1.17}^{+0.97}\\
         M_p\; [M_J] & 0.835\pm0.077 & 0.452_{-0.033}^{+0.034} & 0.273_{-0.018}^{+0.019} & 1.241\pm0.044\\
         M_p\; [M_\oplus] & 265_{-24}^{+25} & 144_{-10}^{+11} & 86.8_{-5.8}^{+5.9} & 395\pm14\\
         T_{eq}\; [K] & 1747\pm27 & 1625_{-21}^{+22} & 1466_{-28}^{+27} & 2183_{-35}^{+36}\\
         a\; [AU] & 0.05014_{-0.00034}^{+0.00035} & 0.54\pm0.0018 & 0.0602\pm0.0024 & 0.0448\pm0.0011\\
         R_p\; [R_J] & 1.5795\pm0.0048 & 1.665_{-0.052}^{+0.052} & 1.19_{-0.04}^{+0.039} & 1.63\pm0.038\\
         R_p\; [R_\oplus] & 17.705_{-0.054}^{+0.053} & 18.66_{-0.57}^{+0.58} & 13.34\pm0.44 & 18.27\pm0.42\\
         \hline
    \end{array}$
\end{table*}

\begin{table*}[t!]
    \centering
    \caption{Estimated physical parameters for HAT-P-67 b, WASP-117 b, WASP-127 b and KELT-18 b}
    \label{tab:tab3}
    $\begin{array}{lcccc}
        \hline
        \hline
         \text{Parameter} & \text{HAT-P-67 b} & \text{WASP-117 b} & \text{WASP-127 b} & \text{KELT-18 b} \\
         \hline
         \text{Transit Parameters} \\
         T_0\; [BJD_{TDB}] & 2458958.08046\pm0.00025 & 2458357.57203\pm0.23 & 2458548.12024\pm0.00019 & 2458932.42707\pm0.00024\\
         P\; [days] & 4.8101088\pm0.000002 & 1.0020593\pm0.0000023 & 4.17806348\pm0.00000072 & 2.8716982\pm0.0000011\\
         b & 0.451_{-0.026}^{+0.023} & 0.424_{-0.058}^{+0.035} & 0.226_{-0.073}^{+0.067} & 0.069_{-0.048}^{+0.054}\\
         R_\star/a & 0.1924_{-0.026}^{+0.024} & 0.0763_{-0.0017}^{+0.0013} & 0.1251_{-0.0016}^{+0.0021} & 0.19387_{-0.00067}^{+0.00089}\\
         R_p/R_\star & 0.08226_{-0.00026}^{+0.00025} & 0.08611_{-0.00076}^{+0.00048} & 0.09971_{-0.00054}^{+0.00051} & 0.08444_{-0.00024}^{+0.00027}\\
         \text{Limb-darkening coefficients}\\
         C_1 & 0.268_{-0.059}^{+0.051} & 0.327_{-0.073}^{+0.053} & 0.3_{-0054}^{+0.045} & 0.235_{-0.052}^{+0.05}\\
         C_2 & 0.129_{-0.078}^{+0.093} & 0.148_{-0.099}^{+0.136} & 0.218_{-0.088}^{+0.109} & 0.149_{-0.076}^{+0.096}\\
         \text{Derived parameters}\\
         T_{14}\; [hr] & 7.023\pm0.02 & 5.846_{-0.018}^{+0.02} & 4.312_{-0.013}^{+0.014} & 4.636_{-0.01}^{+0.011}\\
         a/R_\star & 5.196_{-0.065}^{+0.07} & 13.12_{-0.24}^{+0.25} & 7.99_{-0.13}^{+0.11} & 5.158_{-0.024}^{+0.018}\\
         i\; [deg] & 85.01_{-0.32}^{+0.35} & 88.15_{-0.2}^{+0.22} & 88.38_{-0.52}^{+0.54} & 89.24_{-0.061}^{+0.053}\\
         M_p\; [M_J] & 0.418\pm0.012 & 0.29\pm0.08 & 0.169\pm0.015 & 1.176pm0.036\\
         M_p\; [M_\oplus] & 132.8_{-3.9}^{+3.8} & 92\pm25 & 53.6\pm4.9 & 374_{-12}^{+11}\\
         T_{eq}\; [K] & 1987_{-22}^{+23} & 1179_{-20}^{+21} & 1406\pm23 & 2077\pm38\\
         a\; [AU] & 0.0615\pm0.0022 & 0.0714\pm0.0038 & 0.0495\pm0.0012 & 0.04573\pm0.00086\\
         R_p\; [R_J] & 2.038_{-0.068}^{+0.067} & 0.98_{-0.049}^{+0.05} & 1.293\pm0.027 & 1.568\pm0.029\\
         R_p\; [R_\oplus] & 22.84_{-0.76}^{+0.75} & 10.98_{-0.55}^{+0.56} & 14.5\pm0.3 & 17.57_{-0.33}^{+0.32}\\
         \hline
    \end{array}$
\end{table*}

\begin{table*}[t!]
    \centering
    \caption{Estimated physical parameters for HAT-P-49 b, WASP-62 b, XO-6 b and WASP-34 b}
    \label{tab:tab4}
    $\begin{array}{lcccc}
        \hline
        \hline
         \text{Parameter} & \text{HAT-P-49 b} & \text{WASP-62 b} & \text{XO-6 b} & \text{WASP-34 b} \\
         \hline
         \text{Transit Parameters} \\
         T_0\; [BJD_{TDB}] & 2459422.24182_{-0.0002}^{+0.00019} & 2458326.078712_{-0.000033}^{+0.000035} & 2458817.584438_{-0.000088}^{+0.000087} & 2458546.4222\pm0.00017\\
         P\; [days] & 2.6915575\pm0.0000015 & 4.41193811\pm0.00000015 & 3.7649923_{-0.00000046}^{+0.00000045} & 4.31768509_{-0.00000086}^{+0.00000085}\\
         b & 0.441_{-0.059}^{+0.044} & 0.237_{-0.011}^{+0.013} & 0.7115_{-0.0049}^{+0.0055} & 0.90501_{-0.0099}^{+0.0103}\\
         R_\star/a & 0.2069_{-0.0053}^{+0.0051} & 0.1031_{-0.00025}^{+0.00032} & 0.121_{-0.0007}^{+0.00081} & 0.096_{-0.016}^{+0.019}\\
         R_p/R_\star & 0.08121_{-0.00058}^{+0.00047} & 0.110749_{-0.000099}^{+0.000116} & 0.115267_{-0.00048}^{+0.0005} & 0.1148_{-0.0021}^{+0.0024}\\
         \text{Limb-darkening coefficients}\\
         C_1 & 0.227_{-0.108}^{+0.071} & 0.296\pm0.012 & 0.22_{-0.11}^{+0.12} & 0.128_{-0.092}^{+0.167}\\
         C_2 & 0.17_{-0.11}^{+0.18} & 0.138_{-0.022}^{+0.024} & 0.24_{-0.15}^{+0.14} & 0.16_{-0.11}^{+0.22}\\
         \text{Derived parameters}\\
         T_{14}\; [hr] & 4.25_{-0.021}^{+0.022} & 3.7802_{-0.0024}^{+0.0024} & 3.0055_{-0.0077}^{+0.008} & 2.076_{-0.018}^{+0.023}\\
         a/R_\star & 4.83_{-0.12}^{+0.13} & 9.7_{-0.03}^{+0.023} & 8.265_{-0.055}^{+0.048} & 10.39_{-0.2}^{+0.18}\\
         i\; [deg] & 84.77_{-0.66}^{+0.82} & 88.601_{-0.082}^{+0.069} & 85.061_{-0.07}^{+0.061} & 85_{-0.15}^{+0.13}\\
         M_p\; [M_J] & 1.73\pm0.2 & 0.58\pm0.031 & 4.47\pm0.12 & 0.58_{-0.91}^{+0.9}\\
         M_p\; [M_\oplus] & 550\pm65 & 184.4_{-9.7}^{+9.8} & 1421_{-39}^{+38} & 184.2_{-9.1}^{+9.0}\\
         T_{eq}\; [K] & 2193\pm32 & 1415\pm18 & 1670\pm25 & 1252_{-12}^{+13}\\
         a\; [AU] & 0.0412\pm0.002 & 0.0581_{-0.0013}^{+0.0014} & 0.0742\pm0.0069 & 0.0545_{-0.0039}^{+0.004}\\
         R_p\; [R_J] & 1.448\pm0.061 & 1.39\pm0.032 & 2.17\pm0.2 & 1.264_{-0.091}^{+0.092}\\
         R_p\; [R_\oplus] & 16.23\pm0.068 & 15.58\pm0.36 & 24.3\pm2.2 & 14.2\pm1.0\\
         \hline
    \end{array}$
\end{table*}

\begin{table*}[t!]
    \centering
    \caption{Estimated physical parameters for WASP-77 b, WASP-187 b, KELT-23A b and WASP-101 b}
    \label{tab:tab5}
    $\begin{array}{lcccc}
        \hline
        \hline
         \text{Parameter} & \text{WASP-77 b} & \text{WASP-187 b} & \text{KELT-23A b} & \text{WASP-101 b} \\
         \hline
         \text{Transit Parameters} \\
         T_0\; [BJD_{TDB}] & 2458412.344742\pm0.000044 & 2458769.9841\pm0.0016 & 2458683.911353\pm0.000022 & 2458470.30368\pm0.00018\\
         P\; [days] & 1.3600288\pm0.00000011 & 5.147882\pm0.0000079 & 2.25528756\pm0.000000064 & 3.5857076\pm0.0000011\\
         b & 0.26_{-0.032}^{+0.024} & 0.746_{-0.03}^{+0.021} & 0.5253_{-0.0046}^{+0.0039} & 0.7475_{-0.0116}^{+0.0097}\\
         R_\star/a & 0.1895_{-0.0013}^{+0.0011} & 0.1964_{-0.0087}^{+0.0071} & 0.13151_{-0.00032}^{+0.00029} & 0.1209\pm0.0016\\
         R_p/R_\star & 0.13009_{-0.00039}^{+0.00036} & 0.06412_{-0.00064}^{+0.00062} & 0.13279_{-0.00018}^{+0.0002} & 0.1083_{-0.00065}^{+0.00074}\\
         \text{Limb-darkening coefficients}\\
         C_1 & 0.38_{-0.027}^{+0.024} & 0.2_{-0.14}^{+0.15} & 0.33_{-0.02}^{+0.023} & 0.19\pm0.13\\
         C_2 & 0.085_{-0.049}^{+0.056} & 0.24_{-0.17}^{+0.19} & 0.177_{-0.041}^{+0.033} & 0.28\pm0.16\\
         \text{Derived parameters}\\
         T_{14}\; [hr] & 2.184_{-0.0036}^{+0.0037} & 5.941_{-0.077}^{+0.075} & 2.2861\pm0.002 & 2.727\pm0.013\\
         a/R_\star & 5.277_{-0.03}^{+0.038} & 5.09_{-0.18}^{+0.24} & 7.604_{-0.017}^{+0.018} & 8.27\pm0.11\\
         i\; [deg] & 87.18_{-027}^{+0.37} & 81.57_{-0.56}^{+0.7} & 86.039_{-0.038}^{+0.044} & 84.82_{-0.14}^{+0.15}\\
         M_p\; [M_J] & 1.759_{-0.054}^{+0.053} & 0.801_{-0.083}^{+0.084} & 0.937_{-0.044}^{+0.045} & 0.495_{-0.04}^{+0.041}\\
         M_p\; [M_\oplus] & 559\pm17 & 254_{-26}^{+27} & 298\pm14 & 157\pm13\\
         T_{eq}\; [K] & 1693\pm25 & 1925_{-49}^{+45} & 1513\pm13 & 1573\pm29\\
         a\; [AU] & 0.02345\pm0.00039 & 0.0671_{-0.0027}^{+0.0032} & 0.03523_{-0.00054}^{+0.00053} & 0.0496\pm0.0017\\
         R_p\; [R_J] & 1.209\pm0.019 & 1.766\pm0.036 & 1.287\pm0.019 & 1.36_{-0.042}^{+0.043}\\
         R_p\; [R_\oplus] & 13.55\pm0.21 & 19.79\pm0.4 & 14.43\pm0.22 & 15.24\pm0.48\\
         \hline
    \end{array}$
\end{table*}

\begin{table*}[t!]
    \centering
    \caption{Estimated physical parameters for HAT-P-30 b, HAT-P-8 b, KELT-6 b and HAT-P-17 b}
    \label{tab:tab6}
    $\begin{array}{lcccc}
        \hline
        \hline
         \text{Parameter} & \text{HAT-P-30b} & \text{HAT-P-8 b} & \text{KELT-6 b} & \text{HAT-P-17 b} \\
         \hline
         \text{Transit Parameters} \\
         T_0\; [BJD_{TDB}] & 2458491.91554\pm0.00016 & 2459827.42998_{-0.00024}^{+0.00025} & 2458936.8463\pm0.00065 & 2458719.47536\pm0.0002\\
         P\; [days] & 2.81060078\pm0.00000046 & 3.076318\pm0.000054 & 7.8456067\pm0.0000096 & 1.0338531\pm0.0000023\\
         b & 0.8686_{-0.0047}^{+0.0046} & 0.205_{-0.115}^{+0.097} & 0.34_{-0.18}^{+0.12} & 0.241_{-0.077}^{+0.054}\\
         R_\star/a & 0.15006_{-0.0013}^{+0.0014} & 0.1546_{-0.0024}^{+0.0037} & 0.0948_{-0.0042}^{+0.0051} & 0.04702_{-0.00065}^{+0.00063}\\
         R_p/R_\star & 0.1093\pm0.0012 & 0.09255_{-0.00049}^{+0.00059} & 0.07622_{-0.00094}^{+0.00104} & 0.12003_{-0.00092}^{+0.00097}\\
         \text{Limb-darkening coefficients}\\
         C_1 & 0.21_{-0.15}^{+0.18} & 0.27_{-0.083}^{+0.05} & 0.339_{-0.13}^{+0.073} & 0.375_{-0.059}^{+0.058}\\
         C_2 & 0.28_{-0.19}^{+0.17} & 0.158_{-0.097}^{+0.154} & 0.16_{0.12}^{+0.22} & 0.32\pm0.13\\
         \text{Derived parameters}\\
         T_{14}\; [hr] & 2.246\pm0.014 & 3.919_{-0.016}^{+0.018} & 5.808_{-0.048}^{+0.052} & 4.062_{-0.017}^{+0.024}\\
         a/R_\star & 6.664_{-0.062}^{+0.058} & 6.47_{-0.15}^{+0.1} & 10.55_{-0.54}^{+0.49} & 21.27_{-0.28}^{+0.3}\\
         i\; [deg] & 82.51_{-0.106}^{+0.099} & 88.19_{-0.93}^{+1.03} & 88.14_{-0.77}^{+0.99} & 89.35_{-0.15}^{+0.21}\\
         M_p\; [M_J] & 0.723\pm0.017 & 1.354_{-0.034}^{+0.035} & 0.442\pm0.019 & 0.571\pm0.018\\
         M_p\; [M_\oplus] & 229.7\pm5.5 & 430\pm11 & 140.6\pm6.1 & 181.6\pm5.8\\
         T_{eq}\; [K] & 1712\pm29 & 1726_{-27}^{+29} & 1366_{-31}^{+38} & 804\pm13\\
         a\; [AU] & 0.0412_{-0.035}^{+0.036} & 0.0473\pm0.002 & 0.0747_{-0.0073}^{+0.0075} & 0.0828\pm0.0024\\
         R_p\; [R_J] & 1.414_{-0.035}^{+0.036} & 1.423_{-0.055}^{+0.054} & 1.13\pm0.1 & 0.978\pm0.026\\
         R_p\; [R_\oplus] & 15.85_{-0.39}^{+0.4} & 14.96\pm0.61 & 12.71\pm1.1 & 10.96\pm0.29\\
         \hline
    \end{array}$
\end{table*}

\begin{table*}[t!]
    \centering
    \caption{Estimated physical parameters for HAT-P-34 b, WASP-54 b, HAT-P-13 b and WASP-13 b}
    \label{tab:tab7}
    $\begin{array}{lcccc}
        \hline
        \hline
         \text{Parameter} & \text{HAT-P-34 b} & \text{WASP-54 b} & \text{HAT-P-13 b} & \text{WASP-13 b} \\
         \hline
         \text{Transit Parameters} \\
         T_0\; [BJD_{TDB}] & 2458686.82646\pm0.00025 & 2458931.23637\pm0.00026 & 2459582.97934\pm0.00038 & 2458870.74355_{-0.00042}^{+0.00041}\\
         P\; [days] & 5.4526482\pm0.0000018 & 3.6935995\pm0.0000018 & 2.916415\pm0.000078 & 4.35323_{-0.00011}^{+0.00012}\\
         b & 0.085_{-0.058}^{+0.08} & 0.464_{-0.053}^{+0.041} & 0.726_{-0.021}^{+0.02} & 0.578_{-0.04}^{+0.034}\\
         R_\star/a & 0.07816_{-0.00045}^{+0.00076} & 0.1602_{-0.0042}^{+0.0037} & 0.1814_{-0.0047}^{+0.0053} & 0.1294_{-0.0039}^{+0.0038}\\
         R_p/R_\star & 0.08097_{-0.00041}^{+0.00044} & 0.08822_{-0.00066}^{0.00059} & 0.08413_{-0.00085}^{+0.00075} & 0.09314_{-0.00078}^{+0.00075}\\
         \text{Limb-darkening coefficients}\\
         C_1 & 0.209_{-0.089}^{+0.09} & 0.309_{-0.09}^{+0.055} & 0.28_{-0.17}^{+0.14} & 0.24_{-0.14}^{+0.12}\\
         C_2 & 0.27_{-0.15}^{+0.17} & 0.124_{-0.084}^{+0.156} & 0.26_{-0.17}^{+0.23} & 0.29_{-0.17}^{+0.2}\\
         \text{Derived parameters}\\
         T_{14}\; [hr] & 3.509_{-0.013}^{+0.014} & 4.479\pm0.02 & 3.296_{-0.027}^{+0.029} & 4.016_{-0.033}^{+0.034}\\
         a/R_\star & 12.795_{-0.123}^{+0.074} & 6.24_{-0.14}^{+0.17} & 5.51_{-0.16}^{+0.15} & 7.73_{-0.22}^{+0.24}\\
         i\; [deg] & 89.62_{-0.37}^{+0.26} & 85.73_{-0.48}^{+0.59} & 82.43_{-0.44}^{+0.41} & 85.71_{-0.39}^{+0.42}\\
         M_p\; [M_J] & 3.64\pm0.22 & 0.633_{-0.02}^{+0.021} & 0.905\pm0.029 & 0.517\pm0.024\\
         M_p\; [M_\oplus] & 1156_{-69}^{+70} & 201.2_{-6.4}^{+6.5} & 287.6\pm9.2 & 164.4\pm7.7\\
         T_{eq}\; [K] & 1288_{-11}^{+12} & 1725\pm35 & 1703\pm35 & 1532_{-24}^{+23}\\
         a\; [AU] & 0.0861_{-0.0066}^{+0.0065} & 0.0531_{-0.0026}^{+0.0027}& 0.045\pm0.0017 & 0.052_{-0.0032}^{+0.0033}\\
         R_p\; [R_J] & 1.142_{-0.087}^{+0.086} & 1.569\pm0.07 & 1.436\pm0.04 & 1.323\pm0.07\\
         R_p\; [R_\oplus] & 12.81_{-0.97}^{+0.96} & 17.58_{-0.79}^{+0.78} & 16.1\pm0.45 & 14.83\pm0.82\\
         \hline
    \end{array}$
\end{table*}

\begin{table*}[t!]
    \centering
    \caption{Estimated physical parameters for WASP-73 b, HAT-P-6 b, HAT-P-7 b and KELT-21 b}
    \label{tab:tab8}
    $\begin{array}{lcccc}
        \hline
        \hline
         \text{Parameter} & \text{WASP-73 b} & \text{HAT-P-6 b} & \text{HAT-P-7 b} & \text{KELT-21 b} \\
         \hline
         \text{Transit Parameters} \\
         T_0\; [BJD_{TDB}] & 2458327.67426_{-0.00043}^{+0.00044} & 2458740.18795_{-0.00034}^{+0.00033} & 2458684.77216_{-0.00016}^{+0.00017} & 2459420.24239\pm0.00022\\
         P\; [days] & 4.087298_{-0.0000019}^{+0.0000018} & 3.8529972\pm0.0000014 & 2.20473825\pm0.00000046 & 3.6127732\pm0.0000037\\
         b & 0.0.539_{-0.056}^{+0.041} & 0.681_{-0.025}^{+0.018} & 0.474_{-0.027}^{+0.022} & 0.358_{-0.066}^{+0.051}\\
         R_\star/a & 0.1964_{-0.007}^{+0.0064} & 0.1376_{-0.0035}^{+0.0033} & 0.2387_{-0.0033}^{+0.0029} & 0.1421\pm0.0032\\
         R_p/R_\star & 0.0582\pm0.00043 & 0.08979_{-0.00063}^{+0.00053} & 0.07748_{-0.00034}^{+0.0003} & 0.0994_{-0.00055}^{+0.00051}\\
         \text{Limb-darkening coefficients}\\
         C_1 & 0.272_{-0.089}^{+0.077} & 0.15_{-0.093}^{+0.121} & 0.272_{-0.049}^{+0.053} & 0.284_{-0.065}^{+0.039}\\
         C_2 & 0.14_{-0.1}^{+0.12} & 0.27_{-0.15}^{+0.13} & 0.158_{-0.082}^{+0.079} & 0.084_{-0.058}^{+0.112}\\
         \text{Derived parameters}\\
         T_{14}\; [hr] & 5.65_{-0.035}^{+0.036} & 3.471_{-0.025}^{+0.026} & 3.951\pm0.01 & 4.0968_{-0.022}^{+0.021}\\
         a/R_\star & 5.091_{-0.16}^{+0.19} & 7.27_{-0.17}^{+0.19} & 4.19_{-0.05}^{+0.059} & 7.04_{-0.15}^{+0.16}\\
         i\; [deg] & 83.93_{-0.68}^{+0.83} & 84.62_{-0.28}^{+0.33} & 83.5_{-0.38}^{+0.46} & 87.09_{-0.49}^{+0.59}\\
         M_p\; [M_J] & 1.887\pm0.054 & 1.108\pm0.04 & 1.79\pm0.034 & 3.884_{-0.049}^{+0.05}\\
         M_p\; [M_\oplus] & 600\pm17 & 352\pm13 & 569\pm11 & 1234\pm16\\
         T_{eq}\; [K] & 1890_{-0.48}^{+0.49} & 1723_{-30}^{+29} & 2199_{-30}^{+31} & 2025_{-30}^{+31}\\
         a\; [AU] & 0.049\pm0.0025 & 0.0494_{-0.0023}^{+0.0024} & 0.03897_{-0.00051}^{+0.00056} & 0.0536\pm0.0016\\
         R_p\; [R_J] & 1.173\pm0.046 & 1.275\pm0.053 & 1.5078\pm0.0095 & 1.584\pm0.034\\
         R_p\; [R_\oplus] & 13.14_{-0.51}^{+0.52} & 14.29\pm0.59 & 16.9\pm0.11 & 17.76\pm0.38\\
         \hline
    \end{array}$
\end{table*}

\section{Data analysis and modelling}\label{sec:sec3}

The sections from the long continuous lightcurves from TESS with full transit observations were sliced to obtain the transit lightcurves, also consisting of an adequate amount (longer two times the transit durations) of out-of-transit baselines on either side of the transit events to allow for accurate modelling of the correlated noise components. To remove the large temporal-scale variations in these lightcurves, the out-of-transit sections were modelled using linear/quadratic polynomials, and the best-fit models determined by least Bayesian Information Criterion \cite[BIC, e.g.][]{neath2012bayesian} were subtracted from the entire transit lightcurves. Such large-scale trends could originate from either long-term variability of the host stars or instrumental systematics and are remnants of the preliminary detrending from the SPOC pipeline. Choosing only the out-of-transit sections of the lightcurves for baseline correction makes sure that the correction functions are not affected by the transit signals.

In order to account for the time uncorrelated fluctuations in the lightcurves, which contribute to the overall uncertainties in modelling, the wavelet denoising \citep{Donoho1994IdealDI, 806084, WaveletDenoise2012, 2019AJ....158...39C, 2021AJ....162...18S, 2021AJ....162..221S} method was used. Unlike other smoothing techniques like binning or Gaussian moving average, wavelet denoising does not remove the high-frequency components from the lightcurves, thus limiting the risk of distorting the transit signatures. For discrete wavelet transforms, the PyWavelets \citep{Lee2019} Python package was used. The Symlet family of wavelets \citep{daubechies1988orthonormal}, which are the least asymmetric modified versions of the Debouchis wavelets \citep{daubechies1992ten, 1995ComPh...9..635R}, were used in the analysis. Only a first level of wavelet denoising was performed in order to avoid the risk of oversmoothing the lightcurves, along with the widely accepted Universal Thresholding Law \citep{Donoho1994IdealDI}.

The processed lightcurves were then analysed through a simultaneous modelling of the transit signal along with the correlated noise components. The time-correlated noise terms in the lightcurves can originate from several sources, such as stellar activities and pulsations, as well as various instrumental effects. We have used the Gaussian process (GP) regression \citep{2006gpml.book.....R, 2015ApJ...810L..23J, 2019MNRAS.489.5764P, 2020AA...634A..75B, 2019AJ....158...39C, 2021AJ....162...18S, 2021AJ....162..221S} technique to model these correlated noise components, which is well adopted in several recent studies. The Matern class covariance function with $\nu$ = $3/2$ with two free parameters, i.e. the signal standard deviation and characteristic time scale, was used in the current analysis. Since the GP regression technique is computationally very expensive for large volumes of data, such as in this work, a modified algorithm was adapted to account for this. For each target, a smaller volume of data (usually 3-4 transit lightcurves) was modelled first, keeping the GP regression coefficients as independent variables in order to estimate their best-fit values. These values for the GP regression parameters were then fixed to model all the lightcurves simultaneously. Since the time-correlated noise components in the TESS lightcurves are expected to be similar in relative standard deviation and temporal scale of variation across the time scale of observations for any particular source, this modified algorithm will not affect the transit modelling apart from reducing the computational overheads. The analytical transit formalism given by \citet{2002ApJ...580L.171M} was used to model the transit signals. The Markov-chain Monte Carlo (MCMC) sampling technique was used for this modelling, incorporating the Hastings-Metropolis algorithm \citep{1970Bimka..57...97H}, in which the final sampling consisted of 30 walkers and 5000 iterations for each target.

The estimated mid-transit times were modelled for a linear ephemeris to estimate the transit ephemeris parameters, i.e. mid-transit time of the 0th epoch ($T_0$) and the orbital period ($P$). The estimated parameters were also used along with the host stars properties to estimate other derivable parameters for these systems, such as transit duration ($T_{14}$), inclination angle ($i$), etc. The stellar parameters used for these calculations were adopted from \cite{2017AJ....153..136S}, \cite{2023A&A...672A..34G}, \cite{2017A&A...602A.107B}, \cite{2017MNRAS.465.3693H}, \cite{2016A&A...585A.126W}, \cite{2017AJ....153..211Z}, \cite{2014A&A...568A..81L}, \cite{2020A&A...643A..45S}, \cite{2017A&A...599A...3L}, \cite{2014AJ....147...84B}, \cite{2017AJ....153...94C}, \cite{2013A&A...558A.106M}, \cite{2013PASP..125...48M}, \cite{2020MNRAS.499..428S}, \cite{2019AJ....158...78J}, \cite{2014MNRAS.440.1982H}, \cite{2015A&A...581L...6D}, \cite{2014A&A...570A..80T}, \cite{2013A&A...551A..73F}, \cite{2010ApJ...718..575W}, \cite{2016ApJ...821...26B}, \cite{2014A&A...563A.143D}, \cite{2014A&A...570A..54L}, \cite{2018AJ....155..100J} and \cite{2017MNRAS.464..810B}.

\section{Results and discussions}\label{sec:sec4}

The transit lightcurves corresponding to the first follow-up transit observations from TESS for each target have been shown in Figure \ref{fig:fig1} - \ref{fig:fig6}. These figures show the normalised lightcurves from TESS after the baseline correction, the lightcurves after wavelet denoising, the best-fit transit model, the best-fit GP regression models and the residuals. It can be noticed from these plots that the wavelet denoising has effectively reduced the fluctuations in the lightcurves, which are generally uncorrelated in time, without oversmoothing them. It can also be seen from these plots that GP regression has very efficiently modelled the time-correlated noise components in the lightcurves, which has added to the precision of the modelling by minimising the residuals. Generally, the effectiveness of GP regression is dependent upon the cadence and SNR of the lightcurves. Since the cadence of all the lightcurves shown here is the same (i.e. 120s), the effectiveness of GP regression can directly be compared between the higher SNR and lower SNR lightcurves, with the efficiency of the technique being higher for the former case.

The physical properties for the targets estimated from the analyses have been tabulated in Tables \ref{tab:tab2} - \ref{tab:tab8}, and the best-fit GP regression model parameters have been tabulated in Table \ref{tab:tab9}. It can be noticed from these tables that the precision of the estimated transit parameters is quite high, which is a direct result of the high SNR TESS lightcurves, the large volume of observational data from TESS, and the critical noise treatment algorithm used in this study. The corner plots showing the posterior distributions of the directly estimated transit parameters from the MCMC modelling of the first three targets have been shown in Figures \ref{fig:figc1} - \ref{fig:figc3}, which show the incredible accuracy of the analyses. I have compared the estimated parameters from this work with those from the literature to showcase the improvements brought up by this study.

First, by comparing it with previous works that have only used ground-based photometric observations, I have found that the precision of the estimated parameters from the present study is much better. The major reason behind this is that ground-based observations are severely affected by atmospheric variability and perturbations, which can give rise to higher uncertainties in the estimated parameters. The volume of data obtained by these studies for each target using the ground-based telescopes, which are affected by factors like day-night cycles and weather effects, etc., is very limited, unlike the large volume of follow-up data available from dedicated space-based facilities like TESS. Also, as a result of these limitations from the ground, several of these previous studies have used partial transit observations, which are extremely difficult to treat for accurate baseline corrections and thus will contribute to higher uncertainties in the estimated parameters. This is also the reason why partial transit observations from TESS were not used in the present analyses, the contributing factor also being the availability of a large volume of high SNR full-transit observations. Finally, the majority of the previous studies have not used any sort of sophisticated techniques to treat the noise components in the lightcurves (e.g. wavelet denoising and GP regression), which has resulted in higher uncertainties in the estimated parameters than that could have been achieved otherwise.

\begin{table}[]
    \centering
    \caption{Best-fit Gaussian-process (GP) regression model parameters}
    \label{tab:tab9}
    $\begin{array}{lcc}
        \hline
        \hline
         \text{Target} & \text{$\mathrm{\alpha}$} & \text{$\mathrm{\tau}$}\\
         \hline
        \text{WASP-79 b} & 0.00024_{-0.000032}^{+0.000035} & 0.0047_{-0.0012}^{+0.002}\\
        \text{WASP-94A b} & 0.000418_{-0.000033}^{+0.000035} & 0.00277_{-0.00031}^{+0.00045}\\
        \text{WASP-131 b} & 0.000352_{-0.000024}^{+0.000026} & 0.00265_{-0.00023}^{+0.00025}\\
        \text{WASP-82 b} & 0.000354_{-0.000025}^{+0.000026} & 0.00365_{-0.00041}^{+0.00043}\\
        \text{HAT-P-67 b} & 0.000328_{-0.000024}^{+0.000028} & 0.00353_{-0.0005}^{+0.00052}\\
        \text{WASP-117 b} & 0.000331_{-0.000023}^{+0.000025} & 0.00285_{-0.00025}^{+0.00029}\\
        \text{WASP-127 b} & 0.000555\pm0.000027 & 0.00241_{-0.00017}^{+0.00019}\\
        \text{KELT-18 b} & 0.000397_{-0.000034}^{+0.000038} & 0.00448_{-0.00067}^{+0.00097}\\
        \text{HAT-P-49 b} & 0.000641_{-0.000042}^{+0.000042} & 0.00304_{-0.00032}^{+0.00036}\\
        \text{WASP-62 b} & 0.000326_{-0.000032}^{+0.000033} & 0.00265_{-0.00033}^{+0.0004}\\
        \text{XO-6 b} & 0.000352_{-0.000045}^{+0.000039} & 0.00328_{-0.00056}^{+0.00077}\\
        \text{WASP-34 b} & 0.000433_{-0.000049}^{+0.000046} & 0.00274_{-0.0004}^{+0.00054}\\
        \text{WASP-77A b} & 0.000398_{-0.000046}^{+0.00004} & 0.00262_{-0.00037}^{+0.00049}\\
        \text{WASP-187 b} & 0.00041_{-0.000025}^{+0.000026} & 0.00286_{-0.00029}^{+0.00033}\\
        \text{KELT-23A b} & 0.000125_{-0.000081}^{+0.000082} & 0.0042_{-0.0016}^{+0.004}\\
        \text{WASP-101 b} & 0.000378_{-0.000035}^{+0.000042} & 0.00325_{-0.00048}^{+0.0006}\\
        \text{HAT-P-30 b} & 0.000389_{-0.000041}^{+0.000044} & 0.00359_{-0.00063}^{+0.00078}\\
        \text{HAT-P-8 b} & 0.00039_{-0.000028}^{+0.000028} & 0.00275_{-0.00029}^{+0.00028}\\
        \text{KELT-6 b} & 0.000403_{-0.000023}^{+0.000024} & 0.00319_{-0.0003}^{+0.00033}\\
        \text{HAT-P-17 b} & 0.000398\pm0.000031 & 0.00289_{-0.00031}^{+0.00037}\\
        \text{HAT-P-34 b} & 0.000401_{-0.000043}^{+0.000045} & 0.00326_{-0.00049}^{+0.00055}\\
        \text{WASP-54 b} & 0.000377_{-0.00003}^{+0.000031} & 0.00342_{-0.00045}^{+0.00055}\\
        \text{HAT-P-13 b} & 0.000346_{-0.000036}^{+0.00035} & 0.0028_{-0.00024}^{+0.00045}\\
        \text{WASP-13 b} & 0.000439_{-0.000038}^{+0.000037} & 0.00286_{-0.00033}^{+0.0004}\\
        \text{WASP-73 b} & 0.000412_{-0.000026}^{+0.000027} & 0.00299_{-0.00031}^{+0.00034}\\
        \text{HAT-P-6 b} & 0.000479_{-0.000034}^{+0.000035} & 0.00333_{-0.00038}^{+0.0005}\\
        \text{HAT-P-7 b} & 0.000442_{-0.000037}^{+0.000038} & 0.00329_{-0.00042}^{+0.00052}\\
        \text{KELT-21 b} & 0.000614_{-0.000048}^{+0.000053} & 0.00317_{-0.00044}^{+0.00046}\\
        \hline
    \end{array}$
\end{table}

Several previous studies have used space-based data, such as from TESS, Spitzer and CHEOPS, to characterize the physical properties of the target planets studied in this work. Table \ref{tab:tab10} lists a comparison of the transit parameters, i.e., $b$, $R_{p}/R_\star$ and $a/R_\star$, from these studies with the present work. Although $R_\star/a$ was the parameter that was directly estimated from the transit modelling in this work, the derived values of $a/R_\star$ were used for the comparison, as it is more widely reported in the previous studies.

\begin{table*}
    \centering
    \caption{Comparison of estimated parameters with the previous studies involving observations from space-based instruments.}
    \label{tab:tab10}
    $\begin{array}{lccccc}
        \hline
        \hline
         \text{Target} & \text{Study} & \text{Instrument} & \text{$b$} & \text{$R_{p}/R_\star$} & \text{$a/R_\star$}\\
         \hline
         \text{WASP-79 b} & \text{This work} & \text{TESS} & 0.5573_{-0.0095}^{+0.0091} & 0.10749\pm0.00025 & 7.139\pm0.047 \\
         & \text{\cite{2021AJ....162..263H}} & \text{TESS} & 0.54\pm0.02 & 0.10715\pm0.00061 & -\\
         & \text{\cite{2022AJ....163..228P}} & \text{TESS} & 0.556^{+0.009}_{-0.010} & 0.1072\pm0.0004 & 7.15\pm0.05\\
         & \text{\cite{2023AcA....73..159M}} & \text{TESS} & - & 0.10766^{+0.00052}_{-0.00056} & 7.148^{+0.074}_{-0.066}\\
         \text{WASP-94A b} & \text{This work} & \text{TESS} & 0.233_{-0.054}^{+0.051} & 0.10561_{-0.00045}^{+0.00048} & 7.173_{-0.089}^{+0.075}\\
         & \text{\cite{2021AJ....162..263H}} & \text{TESS} & 0.287\pm0.078 & 0.10679\pm0.00078 & - \\
         \text{WASP-131 b} & \text{This work} & \text{TESS} & 0.724_{-0.018}^{+0.015} & 0.07998_{-0.00061}^{+0.00048} & 8.45_{-0.18}^{+0.2} \\
         & \text{\cite{2021AJ....162..263H}} & \text{TESS} & 0.72\pm0.031 & 0.08039\pm0.00107 & -\\
         & \text{\cite{2022AJ....163..228P}} & \text{TESS} & 0.721_{-0.028}^{+0.025} & 0.0804\pm0.001 & 8.52_{-0.29}^{+0.3}\\
         \text{WASP-82 b} & \text{This work} & \text{TESS} & 0.112_{-0.075}^{+0.088} & 0.07683\pm0.00034 & 4.429_{-0.054}^{+0.029}\\
         & \text{\cite{2021AJ....162..263H}} & \text{TESS} & 0.14\pm0.09 & 0.07731\pm0.00041 & -\\
         & \text{\cite{2022AJ....163..228P}} & \text{TESS} & 0.136_{-0.09}^{+0.095} & 0.0773\pm0.0004 & 4.42_{-0.07}^{+0.04} \\
         \text{WASP-117 b} & \text{This work} & \text{TESS} & 0.424_{-0.058}^{+0.035} & 0.08611_{-0.00076}^{+0.00048} & 13.12_{-0.24}^{+0.25} \\
         & \text{\cite{2022AJ....163..228P}} & \text{TESS} & 0.435_{-0.052}^{+0.041} & 0.0865_{-0.0007}^{+0.0006} & 16.93_{-0.35}^{+0.39}\\
         \text{WASP-127 b} & \text{This work} & \text{TESS} & 0.226_{-0.073}^{+0.067} & 0.09971_{-0.00054}^{+0.00051} & 7.99_{-0.13}^{+0.11} \\
         & \text{\cite{2021AJ....162..263H}} & \text{TESS} & 0.204\pm0.104 & 0.1008\pm0.00083 & -\\
         \text{KELT-18 b} & \text{This work} & \text{TESS} & 0.069_{-0.048}^{+0.054} & 0.08444_{-0.00024}^{+0.00027} & 5.158_{-0.024}^{+0.018}\\
         & \text{\cite{2020AcA....70..181M}} & \text{TESS} & - & 0.0845\pm0.0005 & 4.36_{-0.09}^{+0.11} \\
         \text{WASP-62 b} & \text{This work} & \text{TESS} & 0.237_{-0.011}^{+0.013} & 0.110749_{-0.000099}^{+0.000116} & 9.7_{-0.03}^{+0.023} \\
         & \text{\cite{2021AJ....162..263H}} & \text{TESS} & 0.283\pm0.031 & 0.11165\pm0.00033 & - \\
         & \text{\cite{2022AJ....163..228P}} & \text{TESS} & 0.228_{-0.016}^{+0.015} & 0.1111\pm0.0001 & 9.72\pm0.03\\
         \text{XO-6 b}  & \text{This work} & \text{TESS} & 0.7115_{-0.0049}^{+0.0055} & 0.115267_{-0.00048}^{+0.0005} & 8.265_{-0.055}^{+0.048}\\
         & \text{\cite{2022AcA....72....1M}} & \text{TESS} & - & 0.10747\pm0.00037 & 8.09\pm0.17\\
         & \text{\cite{2022AJ....163..228P}} & \text{TESS} & 0.722\pm0.007 & 0.1158_{-0.0009}^{+0.001} & 8.17\pm0.07 \\
         \text{WASP-34 b}  & \text{This work} & \text{TESS} & 0.90501_{-0.0099}^{+0.0103} & 0.1148_{-0.0021}^{+0.0024} & 10.39_{-0.2}^{+0.18}\\
         & \text{\cite{2021AJ....162..263H}} & \text{TESS} & 0.898\pm0.052 & 0.11998\pm0.00587 & -\\
         & \text{\cite{2022AJ....163..256M}} & \text{Spitzer} & - & 0.1201\pm0.00112 & 10.688\pm0.029\\
         \text{WASP-77A b} & \text{This work} & \text{TESS} & 0.26_{-0.032}^{+0.024} & 0.13009_{-0.00039}^{+0.00036} & 5.277_{-0.03}^{+0.038} \\
         & \text{\cite{2020A&A...636A..98C}} & \text{TESS} & 0.109_{-0.071}^{+0.089} & 0.13354_{-0.0007}^{+0.00074} & 5.332_{-0.081}^{+0.057}\\
         & \text{\cite{2021AJ....162..263H}} & \text{TESS} & 0.344\pm0.04 & 0.11982\pm0.00073 & -\\
         \text{WASP-187 b} & \text{This work} & \text{TESS} & 0.746_{-0.03}^{+0.021} & 0.06412_{-0.00064}^{+0.00062} & 5.09_{-0.18}^{+0.24} \\
         & \text{\cite{2020MNRAS.499..428S}} & \text{TESS} & 0.76\pm0.02 & 0.0591\pm0.0024 & -\\
         \text{KELT-23A b} & \text{This work} & \text{TESS} & 0.5253_{-0.0046}^{+0.0039} & 0.13279_{-0.00018}^{+0.0002} & 7.604_{-0.017}^{+0.018}\\
         & \text{\cite{2020AcA....70..181M}} & \text{TESS} & - & 0.1320\pm0.0006 & 7.556_{-0.045}^{+0.041} \\
         \text{WASP-101 b} & \text{This work} & \text{TESS} & 0.7475_{-0.0116}^{+0.0097} & 0.1083_{-0.00065}^{+0.00074} & 8.27\pm0.11\\
         & \text{\cite{2021AJ....162..263H}} & \text{TESS} & 0.719\pm0.025 & 0.10925\pm0.00147 & -\\
         & \text{\cite{2022AJ....163..228P}} & \text{TESS} & 0.745\pm0.011 & 0.1089_{-0.0007}^{+0.0009} & 8.31\pm0.11 \\
         \text{HAT-P-30 b} & \text{This work} & \text{TESS} & 0.8686_{-0.0047}^{+0.0046} & 0.1093\pm0.0012 & 6.664_{-0.062}^{+0.058}\\
         & \text{\cite{2021AJ....162..263H}} & \text{TESS} &  0.852\pm0.022 & 0.10962\pm0.00258 & - \\
         & \text{\cite{2022AJ....163..208B}} & \text{TESS} & 0.872\pm0.003 & 0.11136\pm0.00064 & - \\
         & \text{\cite{2022MNRAS.513.3444B}} & \text{TESS} & 0.8651\pm0.0029 & 0.10949\pm0.00046 & 6.895\pm0.056 \\
         \text{HAT-P-34 b} & \text{This work} & \text{TESS} & 0.085_{-0.058}^{+0.08} & 0.08097_{-0.00041}^{+0.00044} & 12.795_{-0.123}^{+0.074} \\
         & \text{\cite{2022AJ....163..228P}} & \text{TESS} & 0.126_{-0.086}^{+0.111} & 0.0818\pm0.0007 & 9.86_{-0.2}^{+0.11}\\
         \text{WASP-13 b} & \text{This work} & \text{TESS} & 0.578_{-0.04}^{+0.034} & 0.09314_{-0.00078}^{+0.00075} & 7.73_{-0.22}^{+0.24}\\
         & \text{\cite{2022AJ....163..228P}} & \text{TESS} & 0.574_{-0.042}^{+0.033} & 0.0932_{-0.0011}^{+0.001} & 7.75_{-0.2}^{+0.23}\\
         \text{WASP-73 b} & \text{This work} & \text{TESS} & 0.0.539_{-0.056}^{+0.041} & 0.0582\pm0.00043 & 5.091_{-0.16}^{+0.19}\\
         & \text{\cite{2021AJ....162..263H}} & \text{TESS} & 0.337\pm0.16 & 0.05713\pm0.00083 & -\\
         & \text{\cite{2022AJ....163..228P}} & \text{TESS} & 0.499_{-0.08}^{+0.055} & 0.058_{-0.0006}^{+0.0005} & 5.24_{-0.19}^{+0.23} \\
         \text{HAT-P-7 b} & \text{This work} & \text{TESS} & 0.474_{-0.027}^{+0.022} & 0.07748_{-0.00034}^{+0.0003} & 4.19_{-0.05}^{+0.059}\\
         & \text{\cite{2010ApJ...713L.145W}} & \text{Kepler} & - & 0.0778\pm0.0003 & -\\
         \text{KELT-21 b} & \text{This work} & \text{TESS} & 0.358_{-0.066}^{+0.051} & 0.0994_{-0.00055}^{+0.00051} & 7.04_{-0.15}^{+0.16}\\
         & \text{\cite{2022MNRAS.513.2822G}} & \text{CHEOPS} & 0.4044\pm0.0095 & 0.0987\pm0.0011 & 6.885\pm0.081\\
         \hline
    \end{array}$
\end{table*}

Compared to \cite{2023ApJS..268....2S} (see Table 10 in that paper), where the targets were around brighter sources, it is found that much fewer studies involving space-based instruments larger than TESS have been performed for the targets in the present study. The only three such studies are \cite{2022AJ....163..256M}, \cite{2022MNRAS.513.2822G} and \cite{2010ApJ...713L.145W}, who have studied the properties of WASP-34 b, KELT-21 b and HAT-P-7 b using the observations from Spitzer, CHEOPS and Kepler, respectively. Out of these, only the Spitzer and CHEOPS observations are intended follow-ups, which could be a result of the bias in interest for brighter sources for follow-up studies using more sophisticated instruments, which also emphasizes the importance of the TESS follow-up observations for a large number of known sources with transiting planets. Compared with \cite{2022AJ....163..256M}, the estimated values for $R_{p}/R_\star$ and $a/R_\star$ from the present study are less precise. However, it must be noted that \cite{2022AJ....163..256M} have kept the values for $i$, $P$ and the limb darkening coefficients ($C_1$ and $C_2$) constant in their modelling, which could have contributed to their apparent highly precise estimation of other physical properties. Compared with \cite{2022MNRAS.513.2822G}, the estimated values for $b$ and $a/R_\star$ are less precise, but that for $R_{p}/R_\star$ is more precise in the present work. Compared with \cite{2010ApJ...713L.145W}, the precision in the estimated values of $R_{p}/R_\star$ is similar in the present work. Since $b$ and $a/R_\star$ are not provided in the previous work, compared with the estimated value of $i$ ($83.1\pm0.5$), the precision from the present study is better.

\begin{figure}
	\centering
	\includegraphics[width=\linewidth]{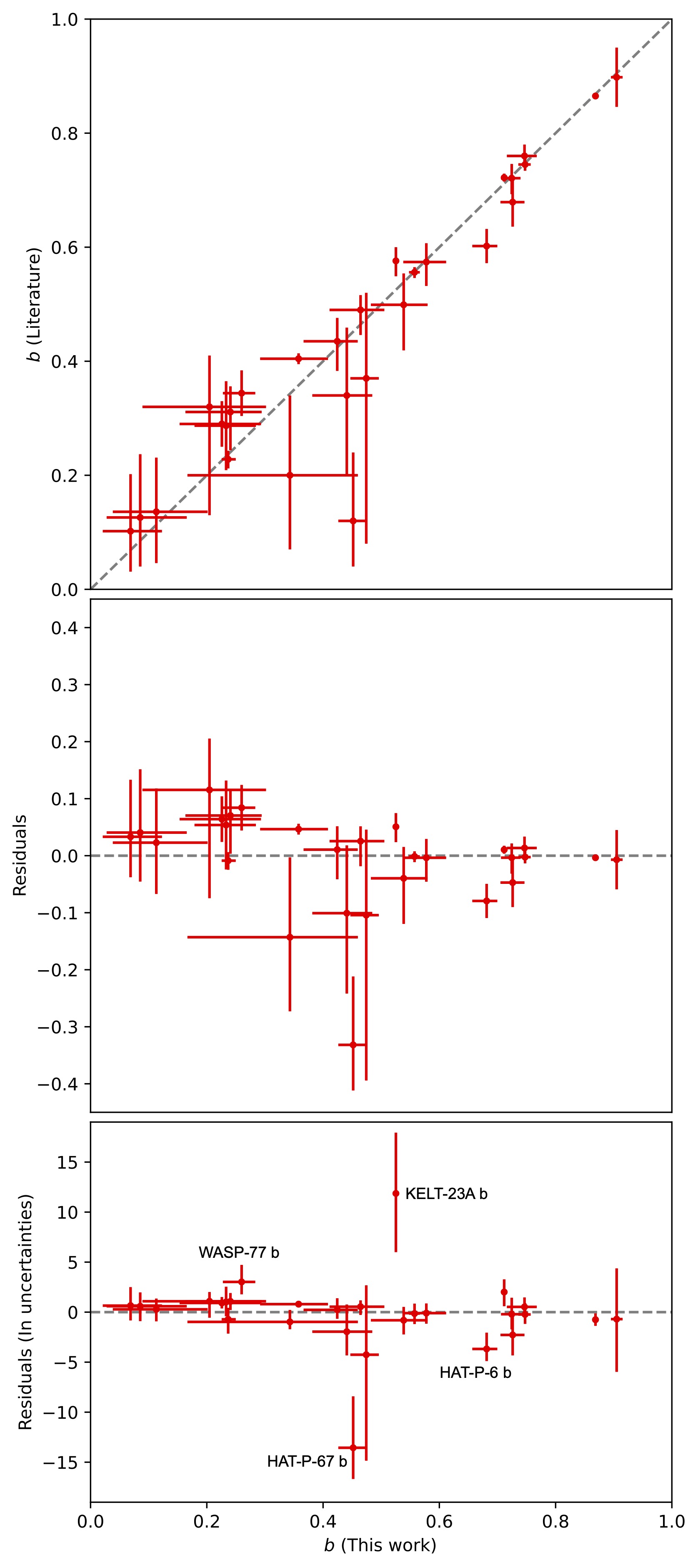}
	\caption{Comparison plot of estimated values of $b$ from this work and the literature (top); the residuals from the comparison in parameter units (middle) and in uncertainties (bottom). The residuals in the units of the uncertainties have been estimated by dividing the residuals by the uncertainties of the estimated parameters from the present work.}
	\label{fig:figp1}
\end{figure}

\begin{figure}
	\centering
	\includegraphics[width=\linewidth]{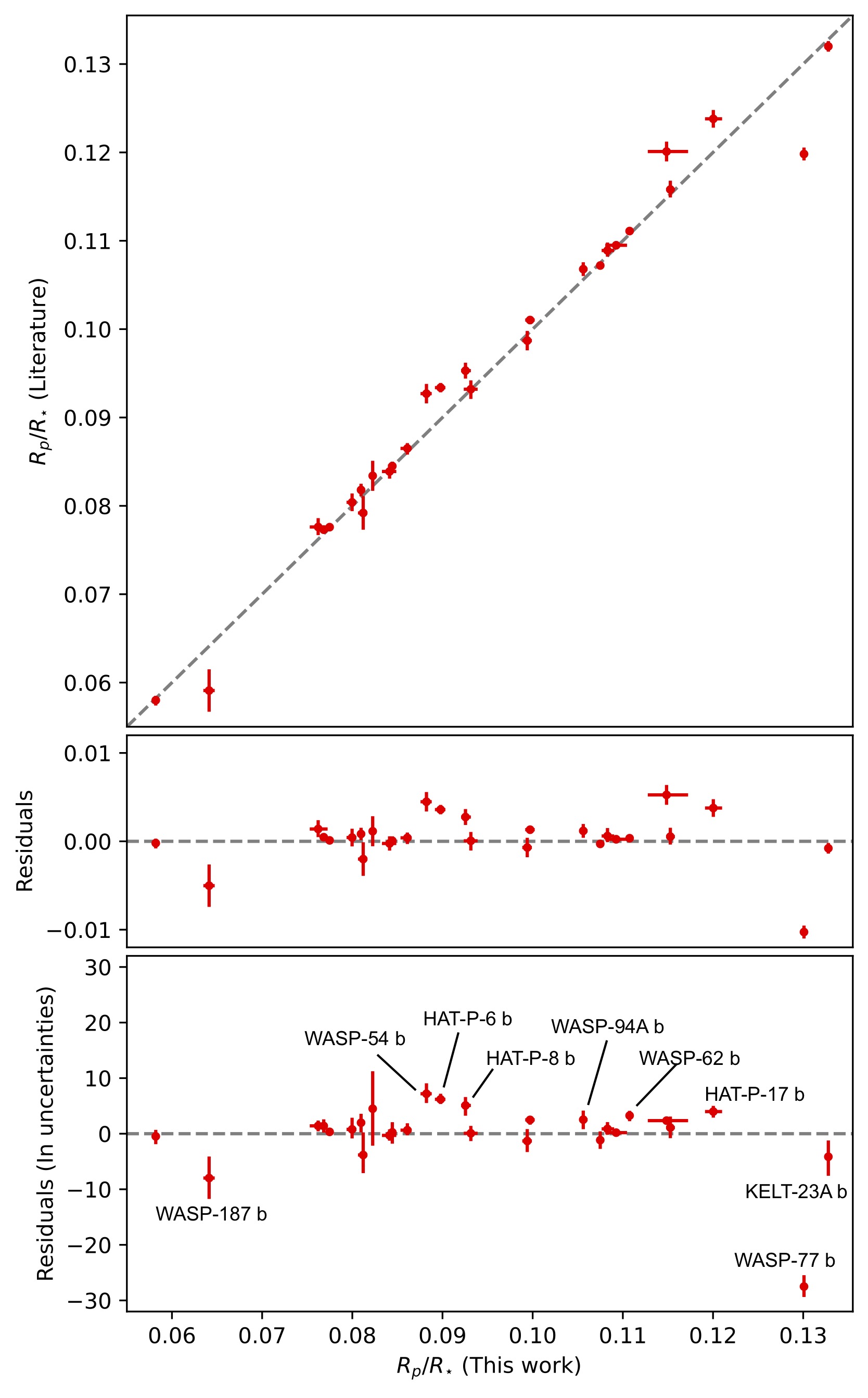}
	\caption{Same as Figure \ref{fig:figp1}, but for $R_p/R_{\star}$.}
	\label{fig:figp2}
\end{figure}

\begin{figure}
	\centering
	\includegraphics[width=\linewidth]{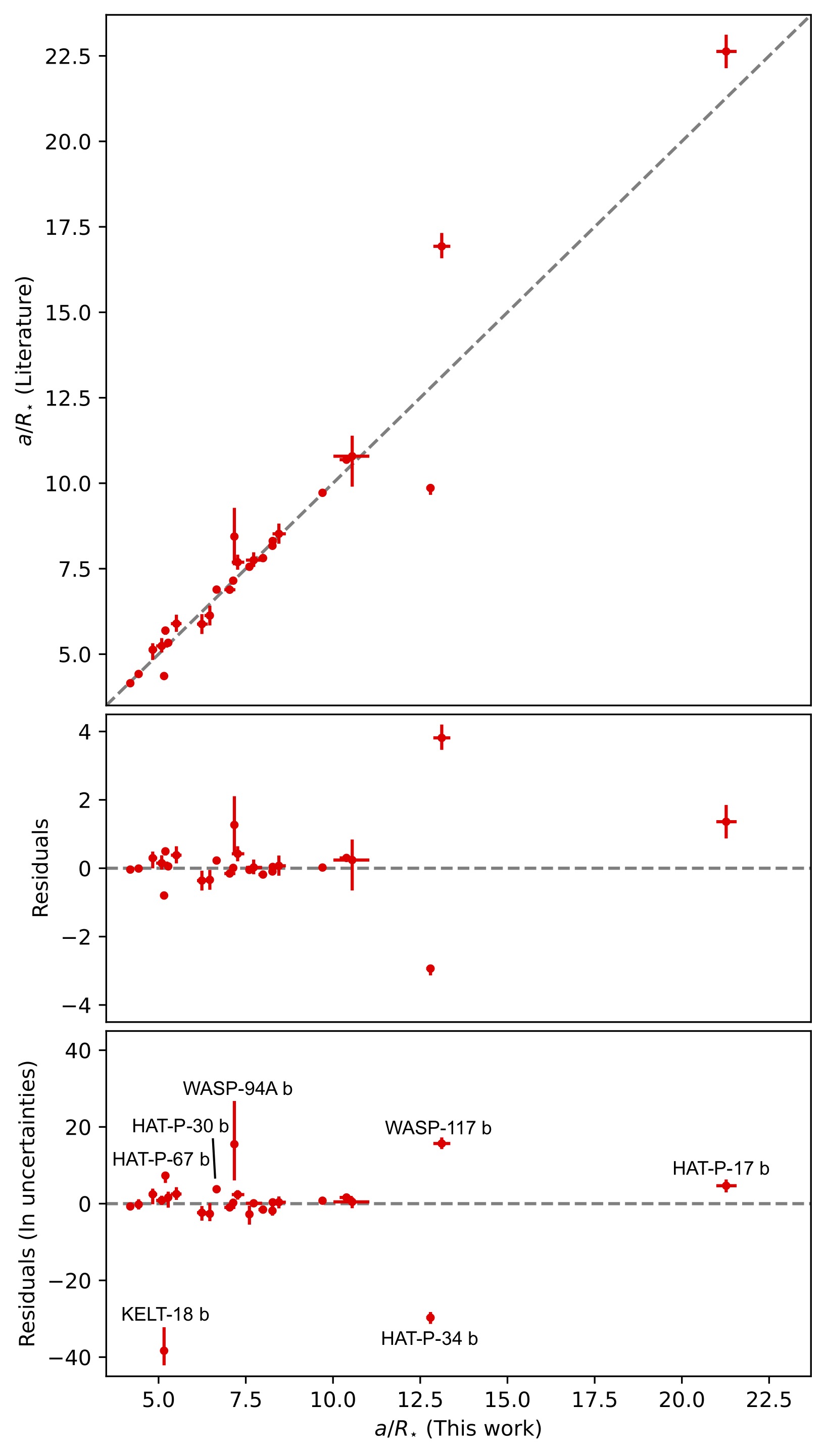}
	\caption{Same as Figure \ref{fig:figp1}, but for $a/R_{\star}$.}
	\label{fig:figp3}
\end{figure}

Comparing with the previous studies involving TESS observations, \cite{2021AJ....162..263H} have studied WASP-79 b, WASP-94A b, WASP-131 b, WASP-82 b, WASP-127 b, WASP-62 b, WASP-34 b, WASP-77A b, WASP-101 b, HAT-P-30 b and WASP-73 b among the targets from this work, with the precision of the estimated parameters from the present study being more precise for all the cases. \cite{2022AJ....163..228P} have studied WASP-79 b, WASP-131 b, WASP-82 b, WSASP117 b, WASP-62 b, XO-6 b, WASP-101 b, HAT-P-34 b, WASP-13 b and WASP-73 b among the targets from this work, with the precision of the estimated parameters from the present study again being more precise for all the cases. Compared with \cite{2023AcA....73..159M}, who has studied WASP-79 b, the estimated parameters from the present work are more precise. Compared with \cite{2020AcA....70..181M}, who has studied KELT-18 b, the estimated parameters from the present work are more precise. For WASP-77A b, the estimated physical properties from the present study are more precise than those from \cite{2020A&A...636A..98C}. Compared with \cite{2020MNRAS.499..428S}, the estimated physical properties for WASP-187 b are more precise in the present work. However, for HAT-P-30 b, the estimated physical properties from the present study are less precise than those reported by \cite{2022AJ....163..208B} and \cite{2022MNRAS.513.3444B}.

One of the major advantages of using a larger volume of follow-up observations for transit analyses is that the estimated physical properties are more accurate statistically, i.e. they are not affected by the inherent bias from various sources in a small volume of datasets. Thus, comparing the estimated physical properties directly with the previous studies can tell us the improvements in accuracy in the estimation of these properties. Figure \ref{fig:figp1} - \ref{fig:figp3} show the comparison plots of the estimated values of $b$, $R_{p}/R_\star$ and $a/R_\star$ from the present work with those from the literature, i.e. \cite{2022AJ....163..228P}, \cite{2021AJ....162..263H}, \cite{2017AJ....153..211Z}, \cite{2017AJ....153..263M}, \cite{2020AcA....70..181M}, \cite{2014AJ....147...84B}, \cite{2022AJ....163..256M}, \cite{2020A&A...636A..98C}, \cite{2020MNRAS.499..428S}, \cite{2019AJ....158...78J}, \cite{2020AcA....70..181M}, \cite{2022AJ....163..208B}, \cite{2009ApJ...704.1107L}, \cite{2014AJ....147...39C}, \cite{2015A&A...581L...6D}, \cite{2012ApJ...749..134H}, \cite{2013A&A...551A..73F}, \cite{2010ApJ...718..575W}, \cite{2008ApJ...673L..79N}, \cite{2013ApJ...764L..22M} and \cite{2022MNRAS.513.2822G}. It can be noticed from these figures that, for several cases, the present study has produced significant corrections to the known physical properties of the target systems. The comparison of the estimated values of $b$ shows that for a number of cases, the estimated parameters from the present study have made statistically significant improvements compared to the previous studies, such as for WASP-77 b, HAT-P-67 b, KELT-23A b and HAT-P-6 b. Comparison of the estimated values for $R_{p}/R_\star$ also shows significant improvements for WASP-187 b, WASP-54 b, HAT-P-6 b, HAT-P-8 b, WASP-34 b, WASP-94A b, WASP-62 b, HAT-P-17 b, WASP-77 b and KELT-23A b. Similarly, for $a/R_\star$, significant improvements in the accuracy of the estimated values are found for KELT-18 b, HAT-P-67 b, HAT-P-30 b, WASP-94A b, HAT-P-34 b, WASP-117 b and HAT-P-17 b, compared to the previous studies.

The estimated mid-transit times were also used to search for any Transit Timing Variation (TTV) trends in these targets. However, no significant TTV trend has been detected for any of the targets studied in this work, with the standard deviations in the O-C analyses being much lower than the uncertainty levels. The O-C diagrams for transit times of WASP-62 b, XO-6 b, KELT-23A b and HAT-P-7 b are shown in Figure \ref{fig:figo1}, which are the targets with the largest number of detected transits (see Table \ref{tab:tab1}). However, for several of the targets, the number of observed transits is still very low to rule out any sort of long-term variations, and thus will require further investigations with future follow-up observations from TESS as well as upcoming facilities, such as PLATO \citep{2014ExA....38..249R} and ET \citep{2022arXiv220606693G}.

I thank the scientific editor for the valuable suggestions in improving the manuscript. I also thank the anonymous reviewer for the valuable comments and suggestions. I acknowledge Fondo Comité Mixto-ESO Chile ORP 025/2022 to support this research. The Geryon cluster at the Centro de Astro-Ingenieria UC was extensively used for the calculations performed in this paper. BASAL CATA PFB-06, the Anillo ACT-86, FONDEQUIP AIC-57, and QUIMAL 130008 provided funding for several improvements to the Geryon cluster. This paper includes data collected by the TESS mission, which are publicly available from the Mikulski Archive for Space Telescopes (MAST). I acknowledge the use of public TOI Release data from pipelines at the TESS Science Office and at the TESS Science Processing Operations Center. Funding for the TESS mission is provided by NASA’s Science Mission directorate. Support for MAST is provided by the NASA Office of Space Science via grant NNX13AC07G and by other grants and contracts. This research made use of Lightkurve, a Python package for Kepler and TESS data analysis.

\bibliography{ms}{}

\clearpage

\section*{Appendix}

\begin{figure*}[h]
	\centering
	\includegraphics[width=0.8\linewidth]{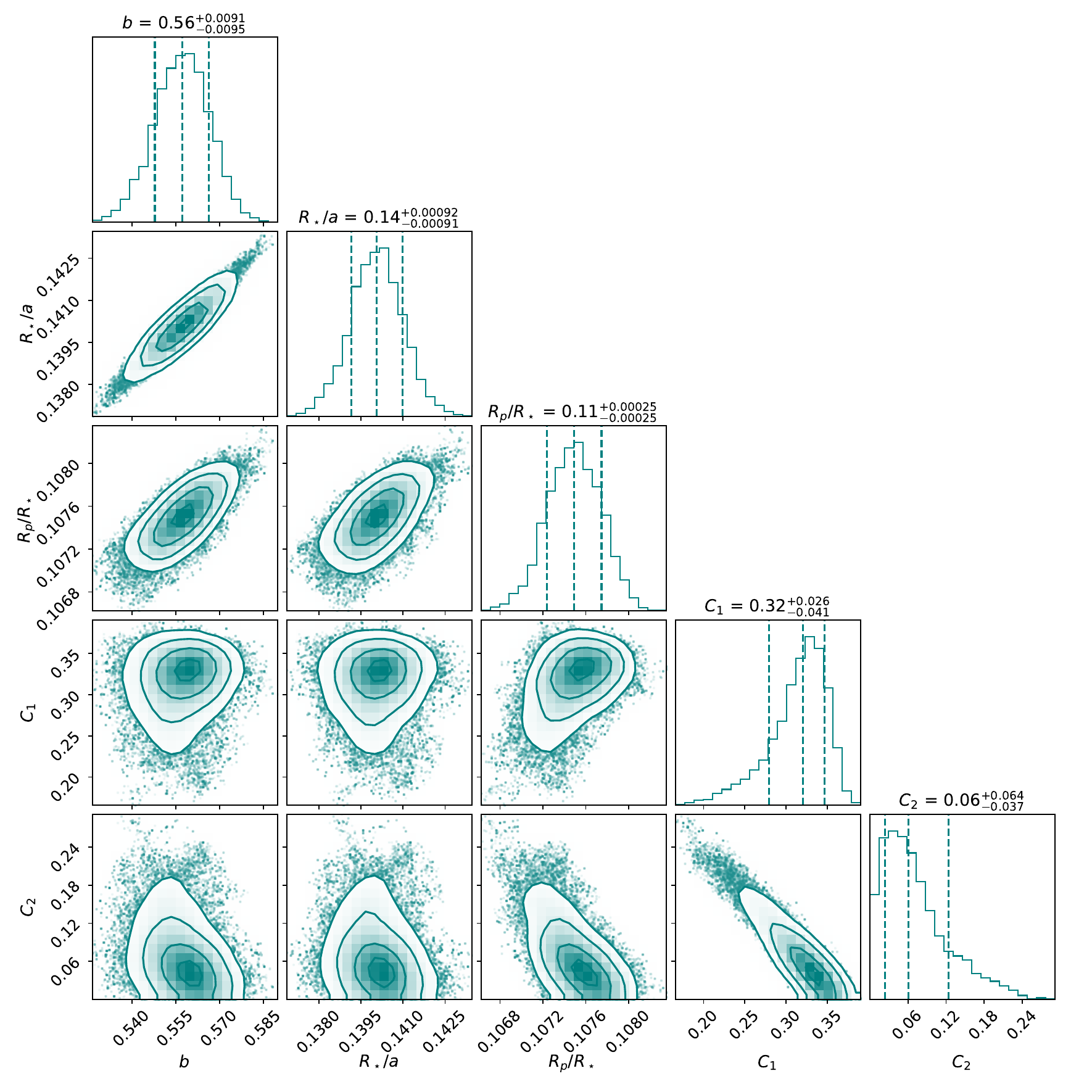}
	\caption{Corner plot showing the posterior distribution of the directly estimated transit parameters from MCMC sampling for WASP-79 b.}
	\label{fig:figc1}
\end{figure*}

\begin{figure*}
	\centering
	\includegraphics[width=0.8\linewidth]{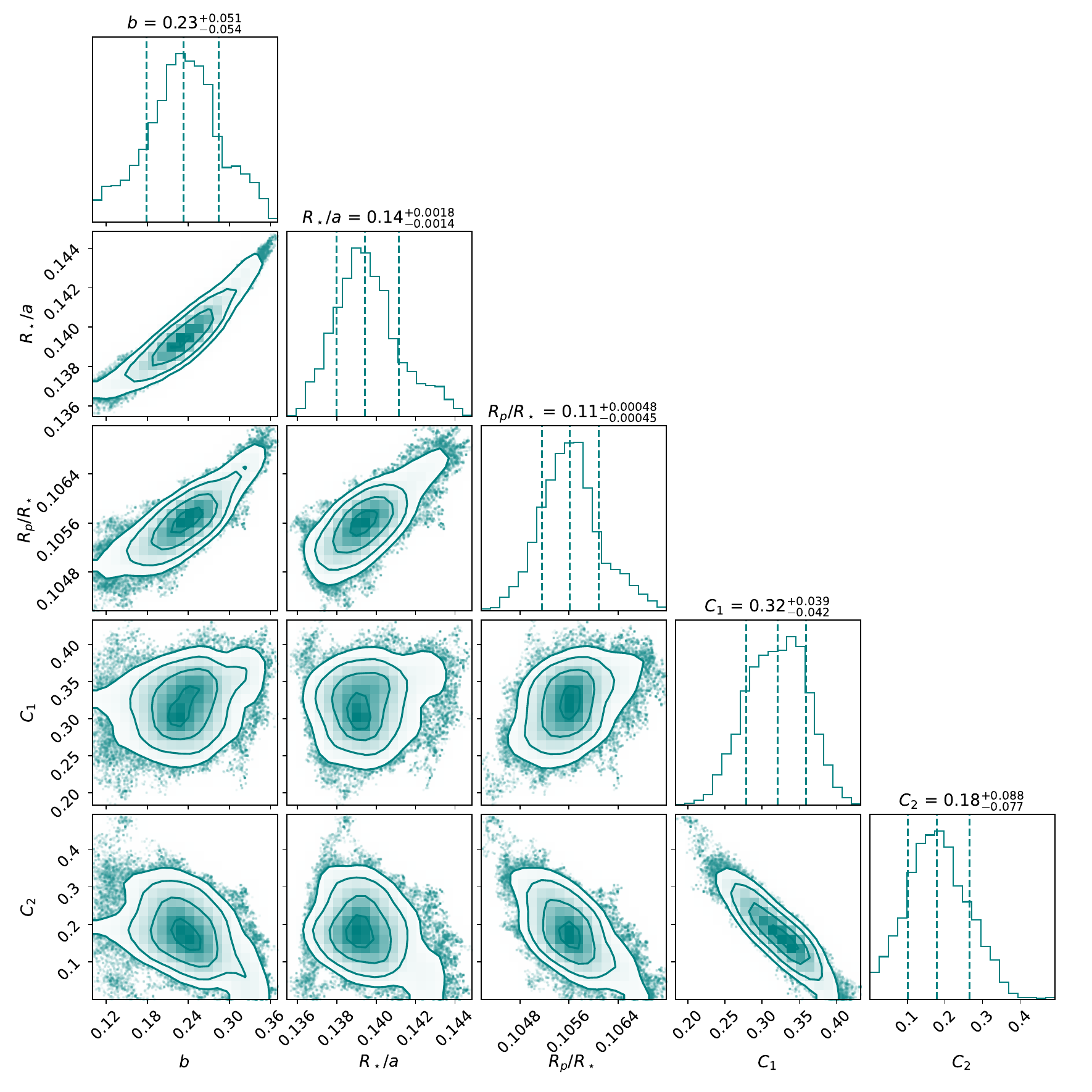}
	\caption{Same as Figure \ref{fig:figc1}, but for WASP-94A b.}
	\label{fig:figc2}
\end{figure*}

\begin{figure*}
	\centering
	\includegraphics[width=0.8\linewidth]{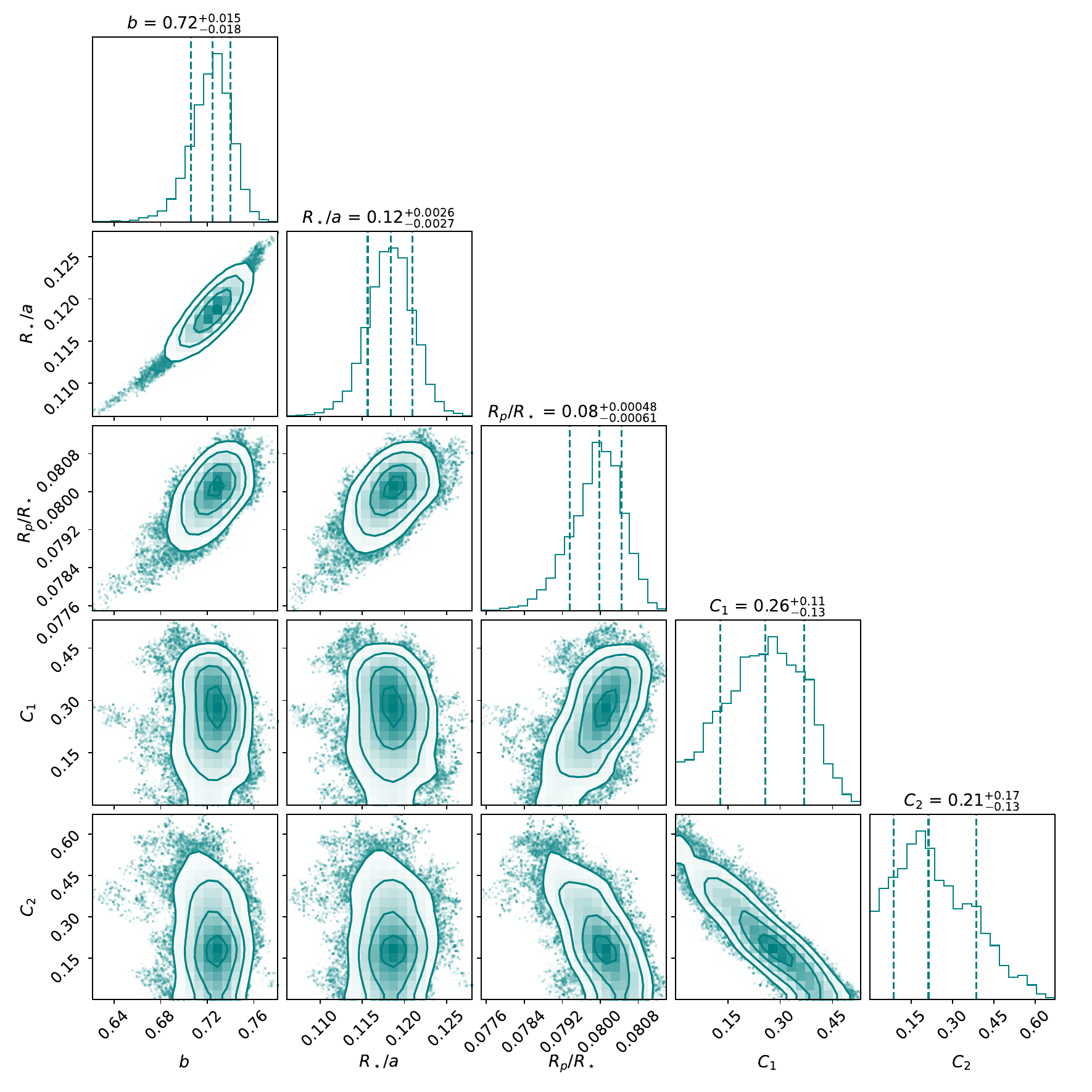}
	\caption{Same as Figure \ref{fig:figc1}, but for WASP-131 b.}
	\label{fig:figc3}
\end{figure*}

\begin{figure*}
	\centering
	\includegraphics[width=0.9\linewidth]{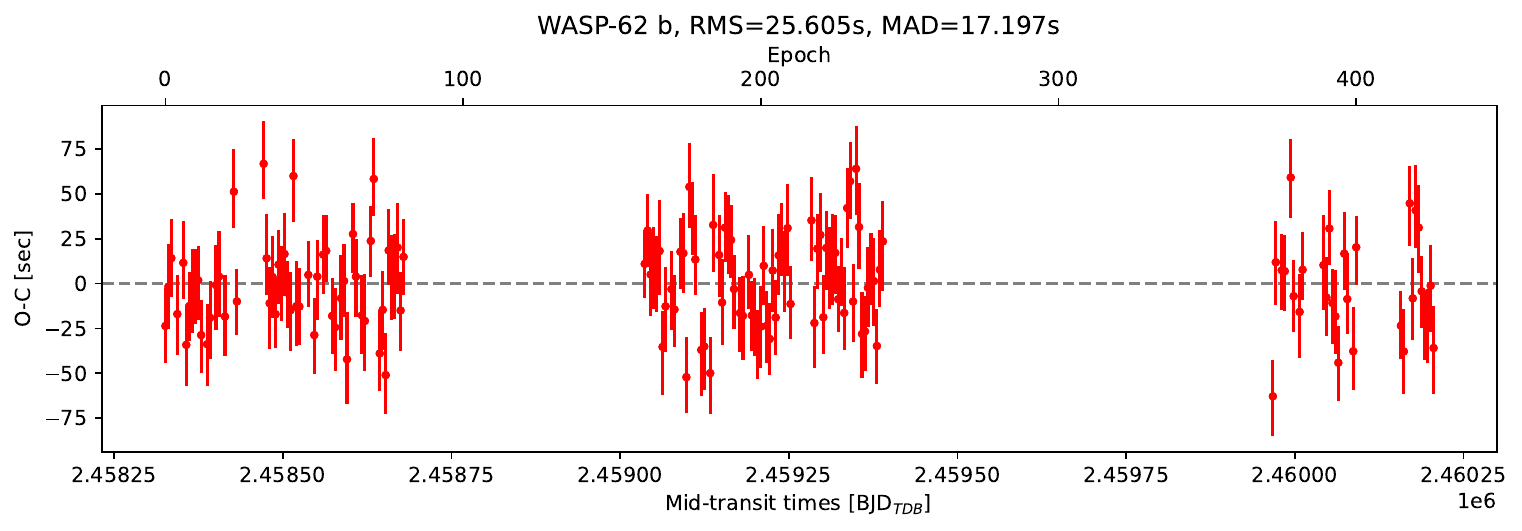}
    \includegraphics[width=0.9\linewidth]{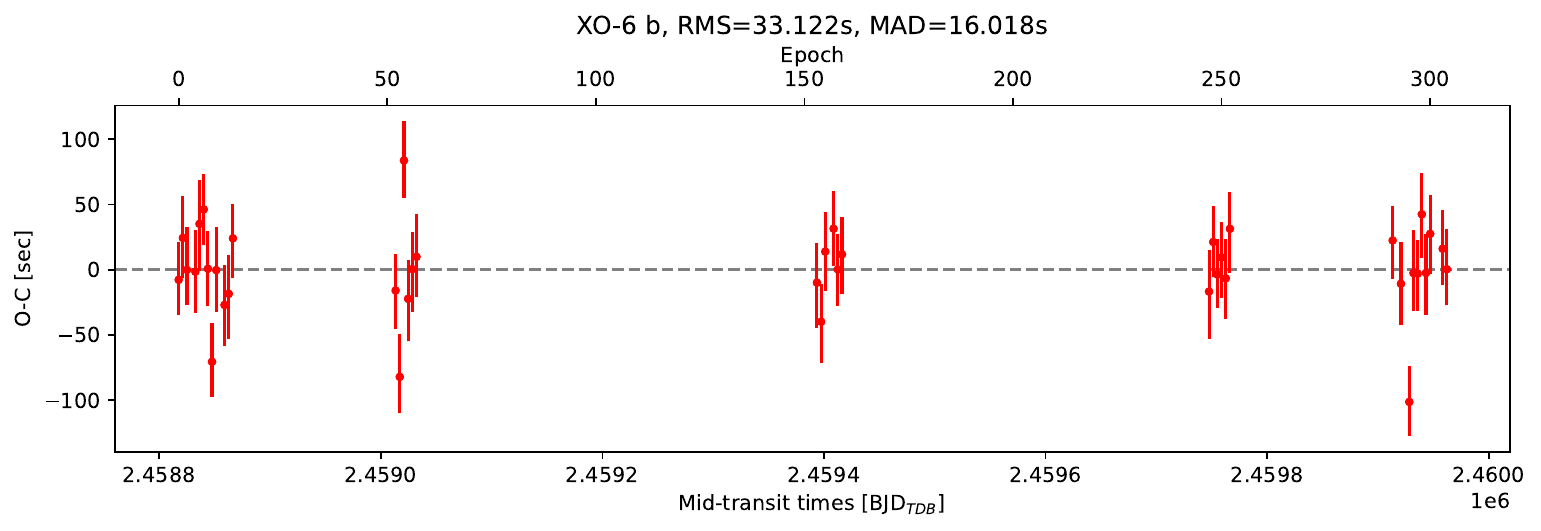}
    \includegraphics[width=0.9\linewidth]{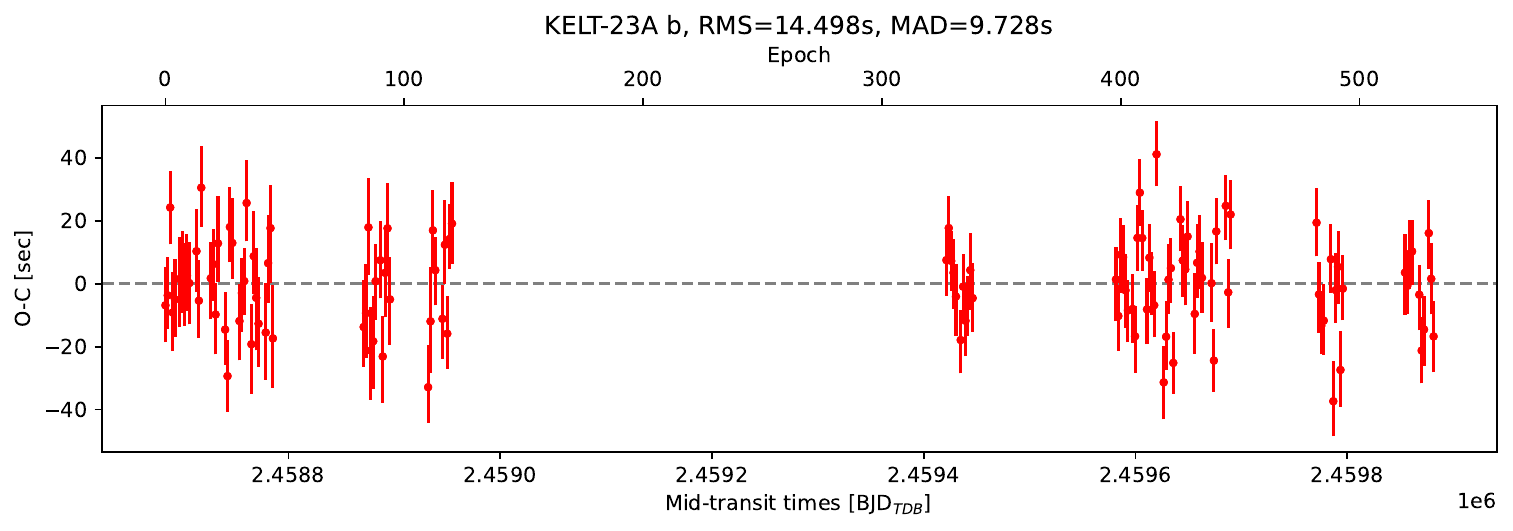}
    \includegraphics[width=0.9\linewidth]{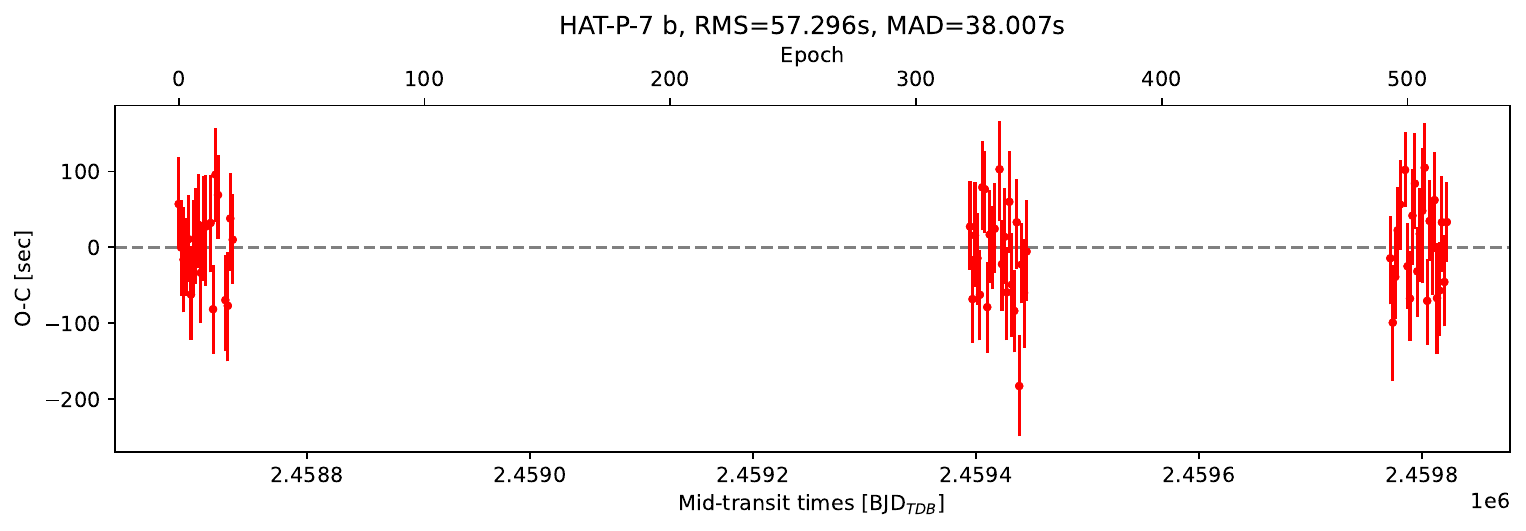}
	\caption{O-C diagrams showing the estimated mid-transit times for WASP-62 b, XO-6 b, KELT-23A b and HAT-P-7 b, considering a linear ephemeris. The Root Mean Square (RMS) and Median Absolute Deviation (MAD) of the mid-transit times have also been indicated.}
	\label{fig:figo1}
\end{figure*}

\end{document}